\documentclass[structabstract]{aa}  
\usepackage{graphicx}
\usepackage{txfonts}
\usepackage{natbib}
\usepackage{multirow}
\usepackage{longtable}
\usepackage{pdflscape}
\usepackage{fixltx2e}
\usepackage[colorinlistoftodos,color=blue!20]{todonotes}

\def\4he{$^4$He}

\def\kms{\mathrm{km\,s}^{-1}}
\def\e#1{\times 10^{#1}}

\def\msol{\mathrm{M}_\odot}

\def\up#1{$^{#1}$}

\def\h2{$\mathrm{H}_2$}
\def\so2{$\mathrm{SO}_2$}
\def\mic{\mathrm{ }\mu\mathrm{m}}
\def\spy{\;\msol~\mathrm{ yr}^{-1}}

\begin{document}
   \title{Water isotopologues in the circumstellar envelopes of M-type AGB stars\thanks{Herschel is an ESA space observatory with science instruments provided by European-led Principal Investigator consortia and with important participation from NASA.}}

   \author{T. Danilovich
          \inst{1}\fnmsep\thanks{Postdoctoral Fellow of the Fund for Scientific Research (FWO), Flanders, Belgium}
          \and
          R. Lombaert\inst{2}
          \and
          L. Decin\inst{1}
          \and
          A. Karakas\inst{3,4}
          \and
          M. Maercker\inst{2}
          \and
          H. Olofsson\inst{2}
          }

   \institute{Department of Physics and Astronomy, Institute of Astronomy, KU Leuven, Celestijnenlaan 200D,  3001 Leuven, Belgium \\
             \email{taissa.danilovich@kuleuven.be}
   \and
   Onsala Space Observatory, Department of Earth and Space Sciences, Chalmers University of Technology, 439 92 Onsala, Sweden  
   \and
   Monash Centre for Astrophysics, School of Physics \& Astronomy, Monash University, VIC, 3800, Australia 
   \and Research School of Astronomy \& Astrophysics, the Australian National University, Canberra ACT 2611, Australia
             }

   \date{Received  / Accepted }

% \abstract{}{}{}{}{} 
% 5 {} token are mandatory
 
  \abstract
   {}
   {In this study we examine rotational emission lines of two isotopologues of water: \h2\up{17}O and \h2\up{18}O. By determining the abundances of these molecules, we aim to use the derived isotopologue --- and hence oxygen isotope --- ratios to put constraints on the masses of a sample of M-type AGB stars that have not been classified as OH/IR stars.}
   {We use detailed radiative transfer analysis based on the accelerated lambda iteration method to model the circumstellar molecular line emission of \h2\up{17}O and \h2\up{18}O for IK~Tau, R~Dor, W~Hya, and R~Cas. The emission lines used to constrain our models come from \textsl{Herschel}/HIFI and \textsl{Herschel}/PACS observations and are all optically thick, meaning that full radiative transfer analysis is the only viable method of estimating molecular abundance ratios.}
   {We find generally low values of the \up{17}O/\up{18}O ratio for our sample, ranging from 0.15 to 0.69. This correlates with relatively low initial masses, in the range $\sim1.0$ to 1.5~$\msol$ for each source, based on stellar evolutionary models. We also find ortho-to-para ratios close to 3, which are expected from warm formation predictions.}
  {The \up{17}O/\up{18}O ratios found for this sample are at the lower end of the range predicted by stellar evolutionary models, indicating that the sample chosen had relatively low initial masses.}
   

   \maketitle
%
%________________________________________________________________

\section{Introduction}

The asymptotic giant branch (AGB) phase follows the main sequence and red giant phase for low- to intermediate-mass stars. AGB stars undergo a period of rapid mass-loss, ejecting matter which forms molecules and condenses into dust in a circumstellar envelope (CSE) around the star. The chemical composition of the CSE depends on the chemical type of the star and AGB stars can be broadly divided into oxygen-rich (M-type) and carbon-rich chemical types, with a third category of S-type stars which have approximately equal abundances of oxygen and carbon.

One of the most abundant molecules found towards M-type stars is \h2O. The abundance and distribution of the main isotopologue, \h2\up{16}O, in AGB stars' CSEs have been studied extensively, for example by \cite{Maercker2008,Maercker2009,Lombaert2013,Khouri2014a,Maercker2016} for M-type stars, \cite{Lombaert2016} for carbon stars, \cite{Schoier2011} and \cite{Danilovich2014} for S-type stars. The millimetre and submillimetre emission of the rarer isotopologues has not been studied in a consistent detailed manner across a sample of stars, although \cite{Decin2010a} previously studied the HIFI isotopologue emission for IK~Tau and \citet{Khouri2014a} performed a detailed analysis for W~Hya.  

A more thorough understanding of the abundances of \h2\up{17}O and \h2\up{18}O will allow us to unravel some of the nucleosynthetic processes that take place during and prior to the AGB phase. For example, first dredge up, which takes place during the red giant branch (RGB) phase, results in an increase in \up{17}O and a decrease in \up{18}O \citep[see][and references therein]{Lattanzio2003,Karakas2014}. The extent of \up{17}O enrichment and \up{18}O depletion depends primarily on the initial mass of the star and does not change appreciably during the second or third dredge ups. Hence, as shown by \cite{De-Nutte2016}, determining the abundances of these isotopes and comparing the results with stellar yields from nucleosynthesis and evolutionary models can allow us to put constraints on the initial masses of the studied AGB stars.

The only mechanism which may significantly change the \up{17}O and \up{18}O abundances after the star has entered the thermally pulsing AGB phase is hot bottom burning (HBB). The onset of HBB, which only takes place in the most massive AGB stars, with masses above $\gtrsim 4~\msol$, rapidly destroys \up{18}O and enhances \up{17}O \citep[see][and references therein]{Lattanzio2003,Karakas2014}. This, of course, has a significant effect on the various ratios involving \up{16}O, \up{17}O, and \up{18}O. It is expected that only AGB stars which are classified as OH/IR stars will be massive enough to have undergone HBB, and evidence of HBB was indeed seen in OH/IR stars by \cite{Justtanont2015}. { The models of \cite{Karakas2016} show that taking the solar \up{17}O/\up{18}O ratio of 0.190 as the initial ratio, the ratio for a $1~\msol$ star will increase to 0.207 after the first dredge up and 0.213 at the first thermal pulse (TP). For stars with higher initial masses the increase in \up{17}O/\up{18}O ratio is more significant: for a $4.5~\msol$ star the ratio will increase to 1.728 after the first dredge up and to 1.781 at the first TP; for an $8~\msol$ star, the ratio will increase to 1.538 after the first dredge up, drop to 0.714 after the second dredge up, due to dredging up \up{18}O from the He shell, and increase significantly to 3943 at the first TP due to the pre-TP onset of HBB.}

Recently, \cite{De-Nutte2016} studied the \up{17}O and \up{18}O abundances based on CO observations in a sample of AGB stars covering all three chemical types. Their determined \up{17}O/\up{18}O ratios spanned $\sim 0.3$ to 2 and indicated initial stellar masses from $\sim 1$ to 1.8~$\msol$, or possibly up to $\sim 4~\msol$, depending on the interpretation of the evolutionary models. 

In this study we look at four M-type AGB stars for which observations of the less common \h2\up{18}O and \h2\up{17}O isotopologues are available from the \textsl{Herschel}/HIFI Guaranteed Time Key Project, HIFISTARS. The high abundance of oxygen relative to carbon in these stars means that \h2\up{16}O is a highly abundant molecule, rivalling or perhaps surpassing the prevalence of CO. Three of the stars (IK~Tau, R~Dor, and R~Cas) in our sample have been previously studied in detail by \cite{Maercker2008,Maercker2009,Maercker2016}, who determined their circumstellar properties from CO observations and determined the \h2\up{16}O abundances. The fourth star, W~Hya, was studied in similar detail by \cite{Khouri2014,Khouri2014a}, who also determined the abundances of \h2\up{18}O and \h2\up{17}O. We include W~Hya in our study to provide a comparison between different modelling methods used to study the molecular envelopes of these stars. There is no overlap of sources between this study and the \cite{De-Nutte2016} study. Although W~Aql and $\chi$~Cyg, for which they study CO isotopologues, were also included in the HIFISTARS programme, no \h2O isotopologues were detected for either S star --- and were not expected to be given the lower abundances of \h2O in those stars --- hence we cannot include them in this study.
We do not expect any of our four stars to have undergone HBB and hence can use the determined abundances and abundance ratios of \h2O isotopologues as gauges of initial stellar mass. A similar study of OH/IR stars, based on the sample presented in \cite{Justtanont2015}, is forthcoming.

%__________________________________________________________________

\section{Sample and observations}\label{obs}

Our sample consists of four M-type AGB stars for which \h2\up{17}O or \h2\up{18}O lines have been detected by the \textsl{Herschel}/HIFI instrument. These stars have a range of mass-loss rates between $10^{-7}$ and $5\e{-6}~\spy$ and have been previously modelled by \cite{Maercker2016}, \cite{Khouri2014,Khouri2014a}, and \cite{Decin2010} to determine mass-loss rates from CO lines and abundances of \h2\up{16}O. 

Some basic information about the four sources is given in Table \ref{radecs}.

\begin{table*}[t]
\caption{Basic information about the four sources in the sample.}\label{radecs}
\begin{center}
\begin{tabular}{ccccccc}
\hline\hline
Star & RA & Dec & Variability &Spectral type & Period & $\dot{M}$\\
 & & & & & [days] & [$\spy$]\\
\hline
IK~Tau & 03 53 28.87 & $+$11 24 21.7& M &M9 & 470 & $5\e{-6}$\\
R~Dor & 04 36 45.59 & $-$62 04 37.8 & SRb&M8e& 332 / 172* & $1.6\e{-7}$\\
W~Hya & 13 49 02.00 & $-$28 22 03.5& M & M7.5-9e&390 & $1\e{-7}$\\
R~Cas & 23 58 24.87 &$+$51 23 19.7& M &	M6.5-9e & 430 & $8\e{-7}$\\
\hline
\end{tabular}
\end{center}
\tablefoot{RA and Dec are given in J2000 co-ordinates. Variability and period information was obtained from the International Variable Star Index (VSX) database. The variability types are M = Mira variable, SRb = semi-regular variable type B. The mass-loss rates, $\dot{M}$, are taken from \cite{Maercker2016}. *Both primary and secondary mode pulsation periods are listed for R~Dor \citep{Bedding1998a}.}
\end{table*}%

\subsection{HIFI data}\label{hifi}

The four stars in our sample, \object{R~Dor}, \object{IK~Tau}, \object{R~Cas}, and \object{W~Hya}, were observed as part of the HIFISTARS Guaranteed Time Key Programme. As part of this programme, the \textsl{Herschel}/HIFI instrument \citep{de-Graauw2010} was used to observe emission lines with high spectral resolution. The full observational results are presented in detail in \citet{Justtanont2012}. However, since those data were published, there have been updates to the main beam efficiencies of the \textsl{Herschel}/HIFI instrument (Mueller et al., 2014\footnote{{\tt http://herschel.esac.esa.int/twiki/pub/Public/Hifi CalibrationWeb/HifiBeamReleaseNote\_Sep2014.pdf}}) and hence we have reprocessed the HIFI data to take this into account \citep[using HIPE\footnote{http://www.cosmos.esa.int/web/herschel/data-processing-overview} version 14.1,][]{Ott2010}.

The detected lines and integrated intensities for all sources are given in Table \ref{hifiobs}.

\begin{table*}[tp]
\caption{\h2O isotopologue observations using HIFI.}\label{hifiobs}
\begin{center}
\begin{tabular}{ccrrrcccc}
\hline\hline
Molecule	&		Transition		&	$\nu\;\;\;$	&	$E_\mathrm{up}$	& $\theta\;$&  IK Tau	&	R Dor	&	W Hya	&	R Cas\\
&&[GHz]&[K]&[\arcsec]&[K $\kms$]&[K $\kms$]&[K $\kms$]&[K $\kms$]\\
\hline
o-\h2$^{17}$O	&	$	3_{1,2}\to 3_{0,3}	$	&	1096.414	&	249	&	19	&	1.97	&	1.46	&	0.56	&	$<$0.49*	\\
	&	$	3_{0,3}\to 2_{1,2}	$	&	1718.119	&	196	&	12	&	$<$2.7*\phantom{$<$}	&	 -	&	1.66	&	-	\\
p-\h2$^{17}$O	&	$	1_{1,1}\to 0_{0,0}	$	&	1107.167	&	53	&	19	&	1.49	&	0.92	&	0.58	&	$<$0.37*	\\
%\hline																			
o-\h2$^{18}$O	&	$	3_{1,2}\to 3_{0,3}	$	&	1095.627	&	249	&	19	&	3.06	&	2.56	&	1.87	&	0.57	\\
p-\h2$^{18}$O	&	$	1_{1,1} \to 0_{0,0}	$	&	1101.698	&	53	&	19	&	4.23	&	2.25	&	1.85	&	0.86	\\
\hline																				
\end{tabular}
\end{center}
\tablefoot{$E_\mathrm{up}$ is the energy of the upper level in the transition, $\theta$ is the half power beam width of the telescope at the corresponding frequency. The integrated line intensity is given for each source and transition. (-) indicates a line not covered by the \textsl{Herschel}/HIFI observations and (*) indicates a non-detection or a marginal detection, primarily used as an upper limit.}
\end{table*}%

\subsection{PACS data}\label{pacs}

The four sources were also observed with \textsl{Herschel}/PACS as part of the MESS Guaranteed Time Key Project \citep{Groenewegen2011}. The PACS spectra cover the 55--100 $\mic$ and 104--190 $\mic$ ranges and the detected lines are not spectrally resolved. As a result, many of the \h2\up{17}O and \h2\up{18}O lines are either known or suspected to be blended with other lines. In many cases this will be evident {through a visual inspection of the lines or} due to there being multiple molecular lines with a central wavelength within the FWHM of the detected PACS line. Such lines are excluded from our analysis, especially since the \h2\up{17}O and \h2\up{18}O are generally relatively faint and any blends with known bright lines such as CO or \h2\up{16}O are not going to provide useful constraints. The line strengths are extracted by fitting a Gaussian line profile. In some cases the FWHM of the fitted Gaussian is more than 20\% larger than the FWHM of the PACS spectral resolution. Such detections are also flagged as blends, even if the secondary components of the blend are unknown, and are also excluded from our modelling. Furthermore, the possibility remains that the lines of interest may be blended with other, unidentified lines. For a detailed description of the data reduction and methodology, see \cite{Lombaert2016}.

The detected PACS lines are listed in Table \ref{pacsobs}. The uncertainties given include both the Gaussian fitting uncertainty and the PACS absolute flux calibration uncertainty of 20\%.

\begin{table*}[tp]
\caption{\h2O isotopologue observations using PACS.}\label{pacsobs}
\begin{center}
\begin{tabular}{ccrrrcccc}
\hline\hline
Molecule	&		Transition		&	$\lambda\;\;\,$	&	$E_\mathrm{up}$	& $\theta\;$&  IK Tau	&	R Dor	&	W Hya	&	R Cas\\
&&[$\mic$]&[K]&[\arcsec]&[$\e{-16}$ W m$^{-2}$]&[$\e{-16}$ W m$^{-2}$]&[$\e{-16}$ W m$^{-2}$]&[$\e{-16}$ W m$^{-2}$]\\
\hline
o-\h2$^{17}$O& $2_{2,1}\to 2_{1,2}$ &180.33&114&13&  blend&x& 0.71 (30\%) & 0.17 (46\%) \\
& $6_{2,5}\to6_{1,6}$ & 94.91&794&9& 3.49 (25\%) & blend &blend &blend\\
%& $8_{3,6}\to8_{2,7}$ & 83.24 &&9& x & x & 3.32 (31\%) & 0.91 (26\%)\\
%& $4_{2,3}\to3_{1,2}$ & 79.16&430&9& blend & blend & blend & 1.04 (35\%)\\
%& $3_{3,0}\to2_{2,1}$ & 66.83&&& 1.56 (28\%) &x&\\
%p-\h2$^{17}$O & $9_{4,6} \to9_{3,7}$&80.69 &&9&1.70 (28\%)&blend&1.92 (43\%)&blend\\
%& $8_{3,5}\to7_{4,4}$ & 80.64&&9& blend & 7.46 (22\%) &3.71 (30\%) & 0.83 (28\%) \\
%& $6_{5,1} \to 6_{4,2}$ & 77.34 &&9& x & 4.59 (27\%) & 5.59 (26\%) & x\\
o-\h2$^{18}$O & $2_{2,1}\to2_{1,2}$& 183.53 & 192&13& 0.91 (29\%) &x &x&x\\
%$\ddagger$& $6_{3,4}\to7_{0,7}$ & 162.32 &&11& 0.52 (30\%)&x&x&x\\
& $2_{2,1}\to 1_{1,0}$& 109.35 &192&10& 4.30 (26\%) & 2.15 (53\%) & x& 0.69 (45\%)\\
& $6_{1,6}\to5_{0,5}$ & 82.44 &641&9& blend&blend&2.25 (25\%)& blend\\
% % & $7_{5,2} \to 7_{4,3}$ & 79.68 &1512&9&1.38 (42\%) & x & x& x\\
%& $7_{2,5} \to 6_{3,4}$ & 74.04 &1123&9&blend & x & x & 0.91 (30\%)\\
&$3_{3,0} \to 2_{2,1}$ & 67.19 &406&9& 2.66 (40\%)& x&x&x\\
& $6_{2,5} \to 5_{1,4}$ & 65.75 &792&9& 1.23 (61\%)& x&x&x\\
& $4_{3,2} \to 3_{2,1}$ & 59.35 &546&9& 2.88 (26\%)& x&x&x\\
p-\h2$^{18}$O & $3_{1,3}\to2_{0,2}$ & 139.59 & 204 &11& 0.99 (38\%) &x &x&x\\
& $3_{2,2}\to2_{1,1}$ & 90.94 & 295 &9& 1.51 (26\%) & x& 1.54 (50\%)&x\\
%& $6_{0,6}\to5_{1,5}$ & 83.59 &&9& x & 5.16 (26\%) & 3.04 (36\%) & 0.84 (27\%)\\
%& $3_{3,1} \to 2_{2,0}$ & 67.88 & 406 &9& 1.32 (26\%) & x&x&x\\
%& $8_{0,8} \to 7_{1,7}$ & 63.72 &&9& x & 11.6 (27\%) & x&x\\
%& $4_{4,0} \to 5_{1,5}$ & 63.51 &&9& 2.78 (28\%)$\ddagger$& x&x&x\\
\hline																				
\end{tabular}
\end{center}
\tablefoot{$E_\mathrm{up}$ is the energy of the upper level in the transition, $\theta$ is the half power beam width of the telescope at the corresponding frequency. The integrated line strengths are given for each source and transition. (x) indicates a non-detection and numbers in brackets indicate percentage errors.}% $\ddagger$ is most likely a misidentification.}
\end{table*}%

\section{Modelling}

\subsection{Established parameters}

The models used to determine the abundances of the \h2O isotopologues in the CSEs of IK~Tau, R~Dor, and R~Cas were based on the circumstellar parameters found by \cite{Maercker2016} as a result of CO line emission modelling. These derived parameters include mass loss rates, gas temperature distributions, dust to gas ratios, and gas expansion velocity profiles. \cite{Maercker2016} also modelled the abundance and distribution of \h2\up{16}O in the same sources, which we have used as a basis for modelling the other isotopologues, assuming that these inhabit the same region around each AGB star as the more common isotopologue. For W~Hya we use the circumstellar properties derived by \cite{Khouri2014,Khouri2014a}, who use a slightly different modelling procedure. Their results were adapted by \cite{Danilovich2016} for SO and \so2 modelling and we use the same method of implementation here. All of these basic CSE parameters are given in Table \ref{stellarprop}.

\begin{figure*}[t]
\sidecaption
\includegraphics[width=0.70\textwidth]{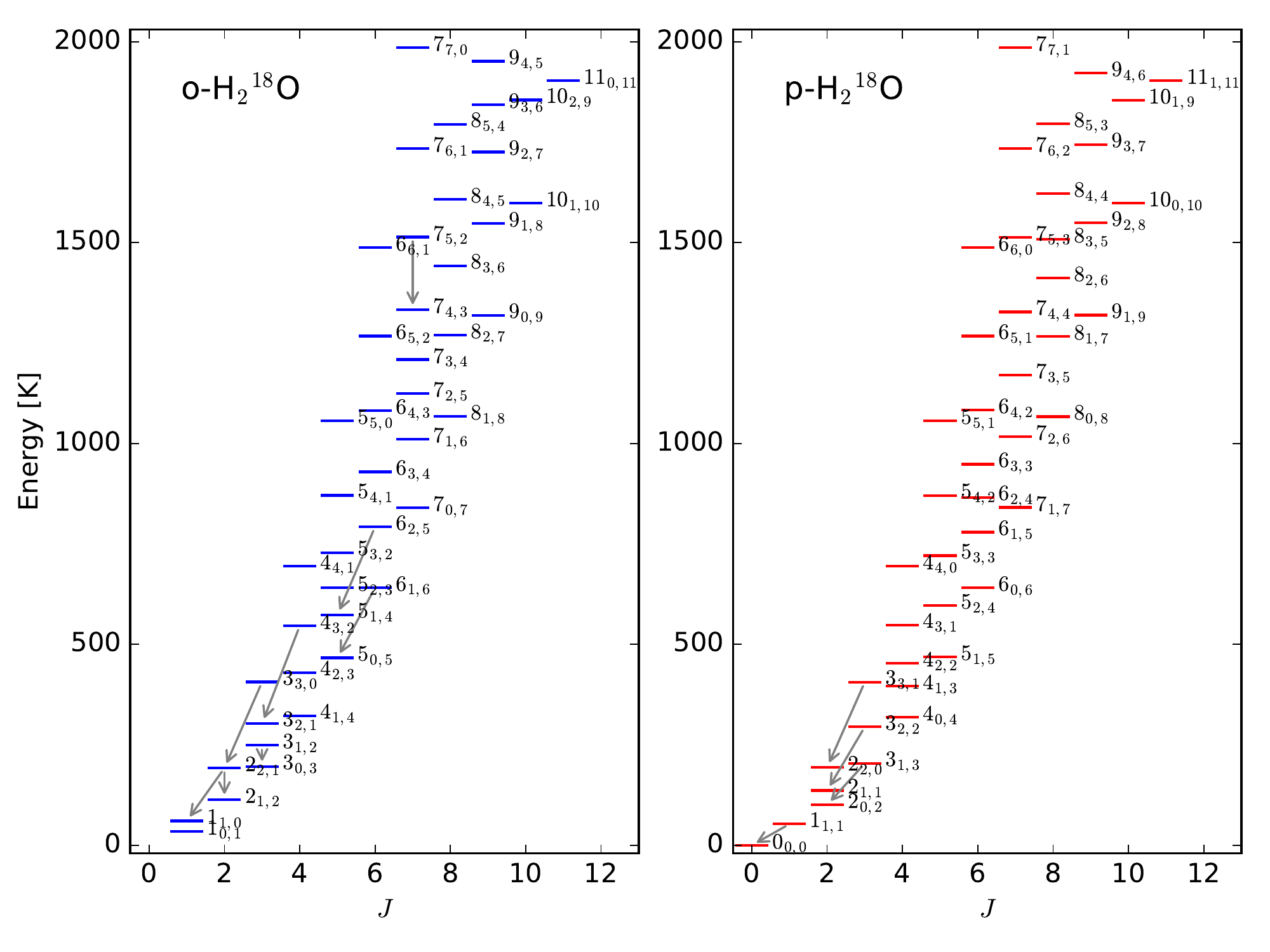}
\caption{An energy level diagram for \h2\up{18}O with ortho energy levels shown on the left in blue and para energy levels shown on the right in red. The quantum numbers are listed to the right of each level in the format $J_{K_a,K_c}$. The grey arrows indicate all the HIFI and PACS transitions used to constrain the models for IK~Tau.}
\label{eld}
\end{figure*}

For the radiative transfer analysis of each isotopologue, the ortho- and para-states of the molecules were treated separately. In each case, we included the lowest 45 rotational energy levels in the ground vibrational state and in the first excited bending mode, $\nu_2=1$. As shown by \cite{Maercker2009} for \h2\up{16}O the $\nu_3=1$ and $\nu_1=1$ vibrationally excited states represent a minimal shift in model predictions when excluded. Our included ground state levels for both ortho and para spin isomers of \h2\up{18}O are shown in Fig. \ref{eld}, along with the transitions observed towards IK~Tau. The equivalently numbered levels were also used for the respective spin isomers of \h2\up{17}O and for \h2\up{16}O by \cite{Maercker2016}. In all cases, the energy levels and radiative rates were obtained from the HITRAN database \citep{Rothman2009}, and the collisional rates used were those for \h2\up{16}O with \h2 from \cite{Faure2007}.

\begin{table}[tp]
\caption{Stellar properties and input from CO models.}\label{stellarprop}
\begin{center}
\begin{tabular}{lcccc}
\hline\hline
	&	IK~Tau	&	R~Dor	&	 W~Hya 	&	 R~Cas 	\\
\hline											
Luminosity [L$_\odot$] 	&	7700	&	6500	&	5400	&	8700	\\
Distance [pc] 	&	265	&	59	&	78	&	176	\\
$\upsilon_\mathrm{LSR}$ [$\kms$] 	&	34	&	7	&	40.5	&	25	\\
$T_*$ [K] 	&	2100	&	2400	&	2500	&	3000	\\
$R_\mathrm{in}$ [$10^{14}$ cm]	&	2.0	&	1.9	&	2.0	&	2.2	\\
$\tau_{10}$ 	&	1.0	&	0.03	&	0.07	&	0.09	\\
$\dot{M}$ [$10^{-7}\spy$] 	&	 $50$ 	&	 $1.6$ 	&	 $1$	&	 $8$ 	\\
$\upsilon_\infty$ [$\kms$] 	&	17.5	&	5.7	&	7.5	&	10.5	\\
$\beta$ 	&	1.5	&	1.5	&	2.0	&	2.5	\\
$R_{\mathrm{H}_2\mathrm{O}}$ [$10^{15}$ cm] & 11 & 1.4 & 1.8 & 3.6\\
\hline
\end{tabular}
\end{center}
\tablefoot{$\upsilon_\mathrm{LSR}$ is the stellar velocity relative to the local standard of rest; $T_*$ is the stellar effective temperature; $R_\mathrm{in}$ is the dust condensation radius, taken to be the inner radius of the model; $\tau_{10}$ is the dust optical depth at $10~\mic$; $\dot{M}$ is the mass-loss rate; $\upsilon_\infty$ is the gas terminal expansion velocity; $\beta$ is the index of the radial velocity profile \citep[see Eq. 1 of][]{Maercker2016}; $R_{\mathrm{H}_2\mathrm{O}}$ is the $e$-folding radius of the Gaussian \h2O radial abundance profile. All parameters are taken from \cite{Maercker2016}.}% \todo[inline]{See Elvire's comments}}
\end{table}%

\subsection{Modelling procedure}

We use an accelerated lambda iteration method code (ALI) to perform detailed radiative transfer modelling of the observed molecular emission lines. ALI has been previously described and implemented by \cite{Maercker2008,Maercker2016}, \cite{Schoier2011}, and \cite{Danilovich2014}. As discussed in more detail in those publications, we assume a smoothly accelerating, spherically symmetric CSE resulting from a constant mass-loss rate. We assume a Gaussian abundance distribution profile with the same $e$-folding radius as \cite{Maercker2016} found for \h2\up{16}O for each source, and vary the central peak abundance to fit the modelled emission lines to the observed emission lines. 

For those molecules where only a single HIFI line is detected, our errors represent the variation in abundance required to produce a variation of 20\% in integrated intensity. For molecules with multiple detections we calculate the best fitting model by minimising a $\chi^2$ statistic, which is defined as
\begin{equation}
\chi^2 = \sum^{N}_{i=1} \frac{\left(I_{\mathrm{mod,}i}-I_{\mathrm{obs,}i}\right)^2}{\sigma_i^2}
\end{equation}
where $I$ is the integrated main beam line intensity for HIFI lines and the flux for PACS lines, $\sigma$ is the uncertainty in the observations (assumed to be 20\% for HIFI lines and listed in Table \ref{pacsobs} for PACS lines), and $N$ is the number of lines being modelled. The errors listed for stars with multiple detections are for a 90\% confidence interval.

\subsection{Refining the observational constraints on the models}\label{refinement}

The largest number of detections were obtained towards IK~Tau, in particular for o-\h2$^{18}$O. In a few cases where the detected PACS line was several orders of magnitude brighter than indicated by model predictions, { a more careful visual inspection of the spectrum resulted in the line being flagged as a blend, generally because the observed and theoretical peaks were significantly misaligned. This, however, does not rule out the possibility of the remaining PACS lines being blended.}
%the line was flagged as a possible blend and excluded from modelling, as discussed in further detail below. Generally, high-$J$ lines ($J\geq 6$) fell into this category. \todo[inline]{Cross-check above}

One unusual case is the o-\h2$^{17}$O ($6_{2,5}\to6_{1,6}$) line at $94.91~\mic$, which was detected towards IK~Tau and fit well with the model. However, it was flagged as a blend { based on a visual inspection towards R~Dor and} based on FWHM towards W~Hya and R~Cas and excluded from modelling. Similarly unusual was the p-\h2$^{18}$O ($6_{0,6}\to5_{1,5}$) line at $83.59~\mic$ which was not detected towards IK~Tau but was detected towards the other three sources. Its non-conformity with the model is less extreme than some apparent blends since the observation is half to two orders of magnitude brighter than the models for the three sources. However, the lack of a detection towards IK~Tau suggests that it is more likely to be a line blend or misidentification in the other sources since there are no other PACS \h2O isotopologue lines which are detected towards other sources but not towards IK~Tau. { Finally, the o-\h2\up{17}O ($4_{2,3}\to 3_{1,2}$) line at 79.16~$\mic$ was flagged as a blend towards R~Dor and W~Hya and visual inspection of the PACS spectra confirms that it is also blended towards IK~Tau and R~Cas. Hence it was removed from modelling for all sources.}

%For IK~Tau, the line that is worst represented by the model is the o-\h2$^{18}$O ($7_{5,2}\to7_{4,3}$) line at $79.68~\mic$. Unlike some of the lines discussed above, this line isn't under-predicted by several orders of magnitude, but only by 80\%. The most likely reason for the discrepancy between model and observation is the fact that the upper energy level is one of the uppermost levels in the ground vibrational state, as can be seen in Fig. \ref{eld}. Hence, there may be some more energetic rotational levels that have a stronger influence on $J_{K_a,K_c} = 7_{5,2}$ than on the lower-$J$ levels, and which are not included in the molecular description of o-\h2O, skewing the model results.

\subsection{Results}\label{sec:results}

The resulting abundances from our radiative transfer calculations, along with some abundance ratios, are given in Table \ref{results}. The HIFI lines plotted with model results are shown in Figures \ref{iktauplots}, \ref{rdorplots}, \ref{whyaplots}, and \ref{rcasplots} for IK~Tau, R~Dor, W~Hya, and R~Cas, respectively. The equivalent plots for the PACS lines are given in Figures \ref{iktaupacs}, \ref{rdorpacs}, \ref{whyapacs}, and \ref{rcaspacs}. In general these show good agreement between observations and models, with only small deviations from the observations. { The most significant of these deviations are the o-\h2\up{17}O ($6_{2,5}\to6_{1,6}$) line for IK~Tau and the p-\h2\up{18}O ($3_{2,2}\to2_{1,1}$) line for W~Hya. Although these are not visibly blended, this is the most likely explanation for the discrepancy, especially when the other lines for those isotopologues are well-represented by our models. The o-\h2\up{18}O ($6_{1,6}\to5_{0,5}$) model line for W~Hya is a slight under-prediction when compared directly with the PACS data. However, as can be seen in Fig. \ref{whyapacs}, the observed line overlaps with the wings of the two lines either side of it, which may contribute some extra flux to our line of interest.} The observed HIFI o-\h2\up{18}O ($3_{1,2}\to3_{0,3}$) line for R~Dor and W~Hya appears to be shifted bluewards in frequency compared with the model. The same is not clearly seen for IK~Tau, but for R~Cas the line has a narrow peak which is also not reproduced by the model. This peak was also seen in some of the SO and \so2 lines towards R~Cas in \cite{Danilovich2016} and could be due to an asymmetric envelope, as found by \cite{Tuthill1994}. Some minor asymmetric features are present in other HIFI lines for all the sources and probably indicate deviations from spherical symmetry in the CSEs. 

We also note that our models indicate that all of the observed lines are optically thick, despite these being rarer isotopologues. Hence, it is not possible to reliably determine abundance ratios by simply comparing line intensities and detailed radiative transfer modelling, as we have performed, is required. { Attempting to determine the \up{17}O/\up{18}O abundance ratios simply by comparing the line intensities gives ratios which differ from those derived through radiative transfer modelling by factors of approximately 2, depending on the choice of line and after taking the difference in Einstein coefficients into account.}

Plots indicating the goodness of fit of the various lines constraining the models are shown in Fig. \ref{fitplots} for those molecules with more than one observed line. These plots show some scatter in how well the models fit the observations, with some of the PACS lines being the worst offenders as discussed in more detail below. However, there are no clear overall trends with energy for over- or under-predicted lines.

\begin{table*}[tp]
\caption{\h2O isotopologue peak fractional abundances with respect to \h2, $f_0$, from model results.}\label{results}
\begin{center}
\begin{tabular}{cccccc}
\hline\hline
& IK Tau	&	R Dor	&	W Hya	&	R Cas\\
\hline
o-\h2$^{17}$O&$(2.9\pm1.0)\e{-7}$& $(3.4\pm0.8)\e{-7}$ & $(3.7\pm0.9)\e{-7}$&$(5.0\pm1.4)\e{-7}$\\
p-\h2$^{17}$O& $(3.8\pm0.9)\e{-8}$& $(4.8\pm1.1)\e{-8}$ &$(6.9\pm1.7)\e{-8}$&$\le3\e{-8}$\\
o-\h2$^{18}$O&$(4.2\pm1.2)\e{-7}$& $(6.3^{+2.7}_{-2.2})\e{-7}$&$(2.5\pm0.9)\e{-6}$ & $(2.8^{+1.8}_{-1.2})\e{-7}$\\
p-\h2$^{18}$O&$(1.5\pm0.4)\e{-7}$& $(1.6\pm0.4)\e{-7}$&$(4.0^{+2.1}_{-1.8})\e{-7}$&$(1.0\pm0.3)\e{-7}$\\
\hline
o-\h2\up{16}O* & $3.5\e{-4}$ & $2\e{-4}$ &$6\e{-4}$ &$6\e{-5}$ \\
p-\h2\up{16}O* & $7\e{-5}$ &$5\e{-5}$& $3\e{-4}$& $2\e{-5}$\\
\hline
o-\h2\up{17}O/o-\h2\up{18}O & $0.69\pm0.31$ & $0.54\pm0.26$ & $0.15\pm0.07$ &$\dagger$\\
p-\h2\up{17}O/p-\h2\up{18}O & $0.25\pm0.09$ & $0.30\pm0.10$ & $0.17\pm0.09$&$\le0.30\pm0.09$\\
o-\h2\up{16}O/o-\h2\up{17}O & 1210 & 588 & 1620 &$\dagger$\\
p-\h2\up{16}O/p-\h2\up{17}O & 1840 & 1040 & 4350 &$\ge667$\\
o-\h2\up{16}O/o-\h2\up{18}O & 833 & 317 & 240 &214\\
p-\h2\up{16}O/p-\h2\up{18}O & 466 & 313 & 750 &200\\
\hline
OPR (\h2\up{17}O) & $7.6\pm3.2$ &$7.1\pm2.3$ & $5.4\pm1.9$ &$\dagger$\\
OPR (\h2\up{18}O) & $2.8\pm1.1$ &$3.9\pm1.8$ & $6.3\pm3.9$ & $2.8\pm2.2$\\
\hline																				
\end{tabular}
\end{center}
\tablefoot{* \h2\up{16}O abundances taken from \cite{Maercker2016} for IK~Tau, R~Dor and R~Cas, and from \cite{Khouri2014a} for W~Hya. $\dagger$ indicates ratios not included due to the inaccurate abundance for o-\h2\up{17}O. See text for details.}
\end{table*}%

\begin{figure}[t]
\begin{center}
\includegraphics[width=0.5\textwidth]{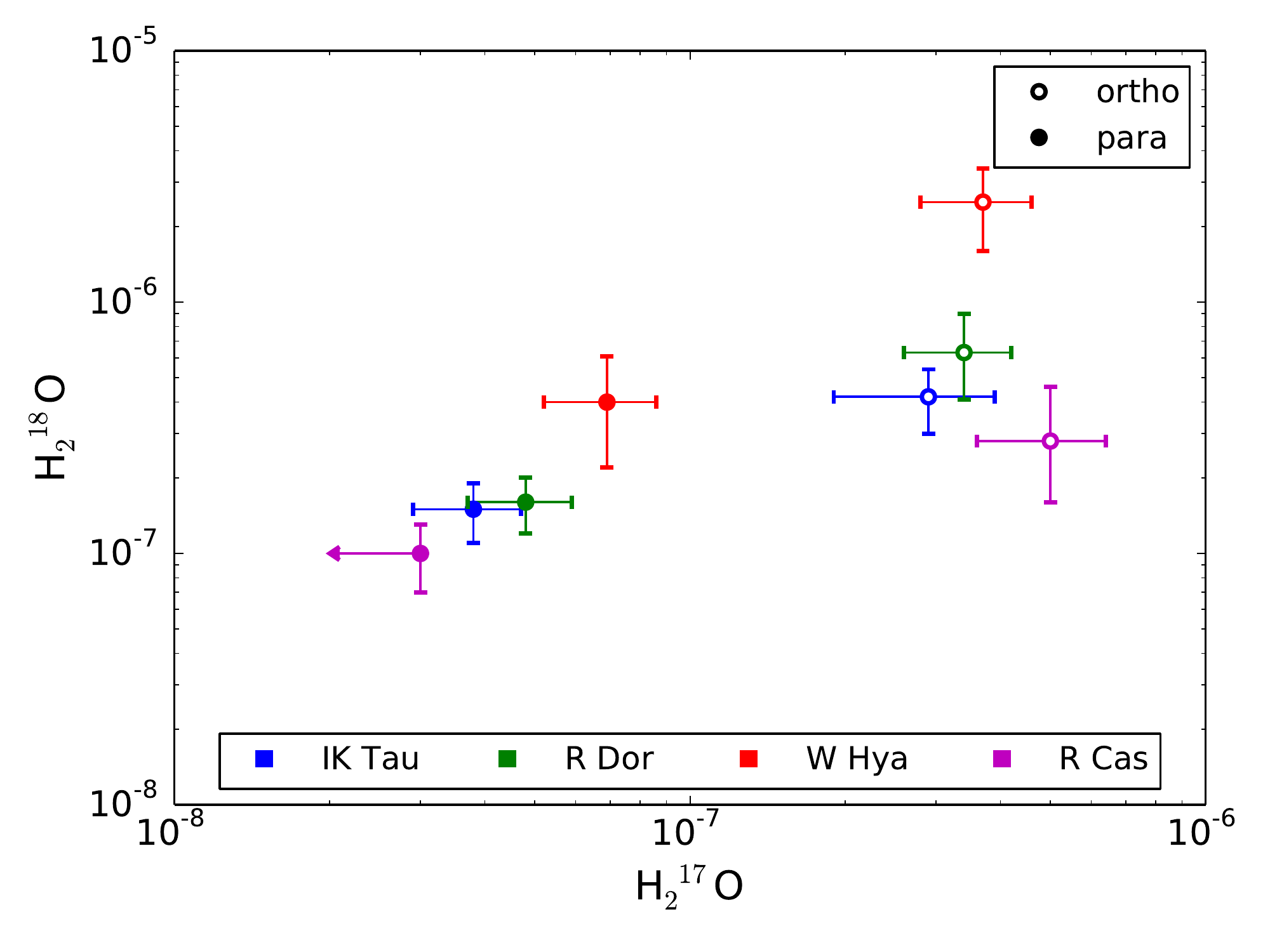}
\caption{A visual representation of the \h2\up{17}O to \h2\up{18}O ratios, separated into ortho (open circles) and para (filled circles) spin isomers and colour-coded by source (see lower legend).}
\label{17to18}
\end{center}
\end{figure}

Overall we found lower abundances for \h2\up{17}O than for \h2\up{18}O. A visual representation of these results is plotted in Fig. \ref{17to18}, showing the stars mostly clustered along a line, with \h2\up{17}O/\h2\up{18}O ratios in the range $\sim0.2$--0.7. The exception to this is R~Cas, which only had non-detections in the HIFI range for \mbox{o-\h2\up{17}O} and one PACS detection.
% which we had difficulty fitting simultaneously. The under-predicted ($4_{2,3}\to3_{1,2}$) line was not as dramatically under-predicted as the excluded lines discussed in Sect. \ref{refinement}, but could possibly be blended, since the same transition towards R~Dor and W~Hya has been flagged as a blend. For the line towards IK~Tau the model only reproduces 43\% of the observed brightness, higher than the R~Cas model, but hardly a good fit. Including this line in our R~Cas modelling, we find a high o-\h2\up{17}O abundance of $(1.3\pm0.6)\e{-6}$. However, if we exclude that line from our model and primarily fit the other detected PACS line (since the HIFI non-detection provides only an upper limit), the modelled abundance reduces by about a factor of 3 to $(5.0\pm1.4)\e{-7}$, bringing it a bit closer in line with the other sources. 
This does not leave us with a very reliable model and in the various plots of results, the R~Cas o-\h2\up{17}O datapoint tends to be an outlier. The abundance ratios involving this result, listed in Table \ref{results}, are not in agreement with the upper limits given by the p-\h2\up{17}O results. 
%Nor do they fall in the same general region as the other three sources nor match up with predictions for possible ratios from evolutionary models (see Sect. \ref{massdet} for more details). 
If we were to accept the unexpectedly high derived abundance for o-\h2\up{17}O, then we would also expect the spectrally resolved HIFI detections of \h2\up{17}O to be brighter than those of \h2\up{18}O. This is not the case --- \h2\up{17}O is not conclusively detected with HIFI --- and so we must conclude that our results for \h2\up{17}O, and especially o-\h2\up{17}O, are unreliable in the case of R~Cas.

\begin{figure*}[t]
\begin{center}
\includegraphics[width=0.49\textwidth]{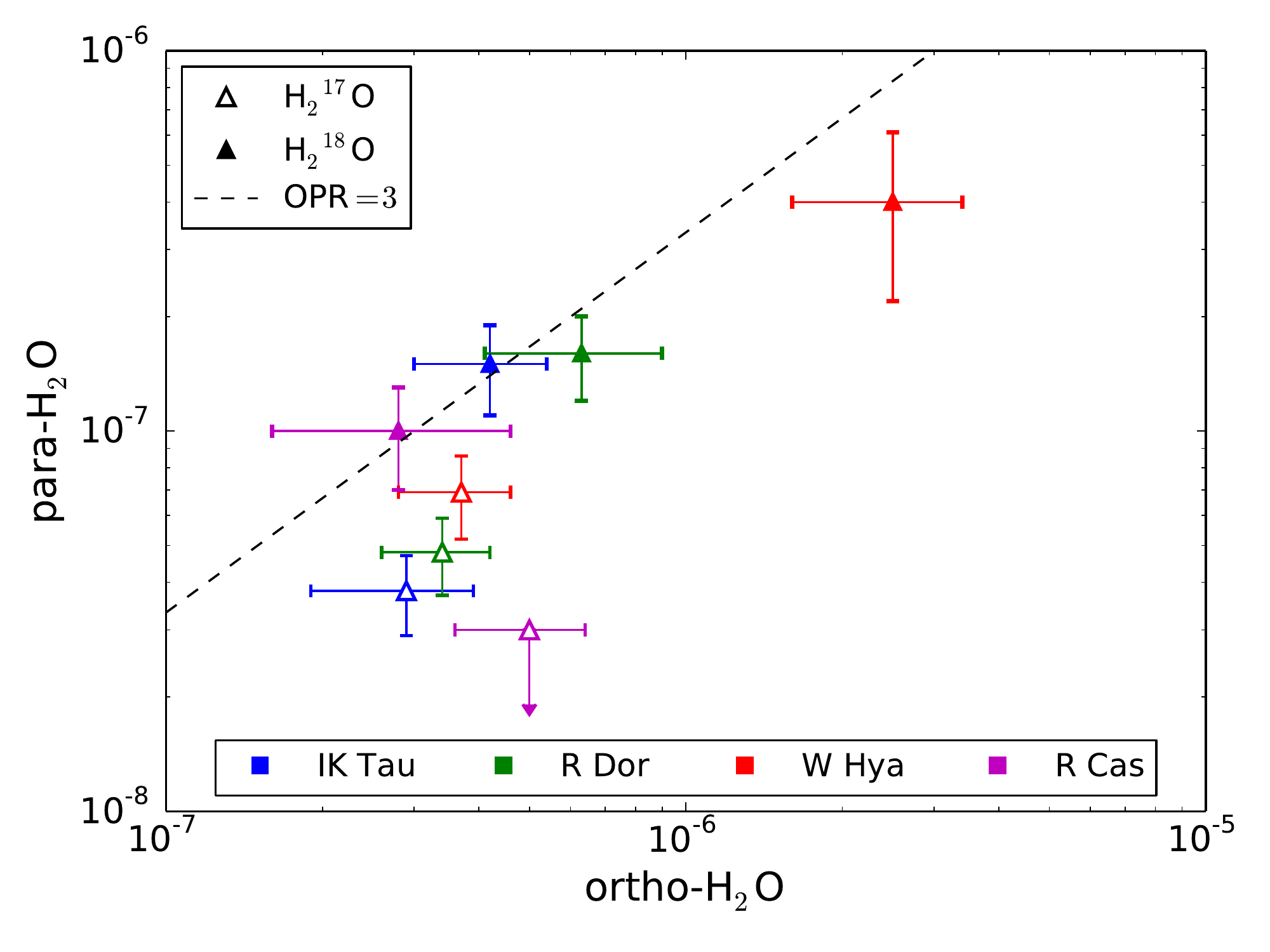}
\includegraphics[width=0.49\textwidth]{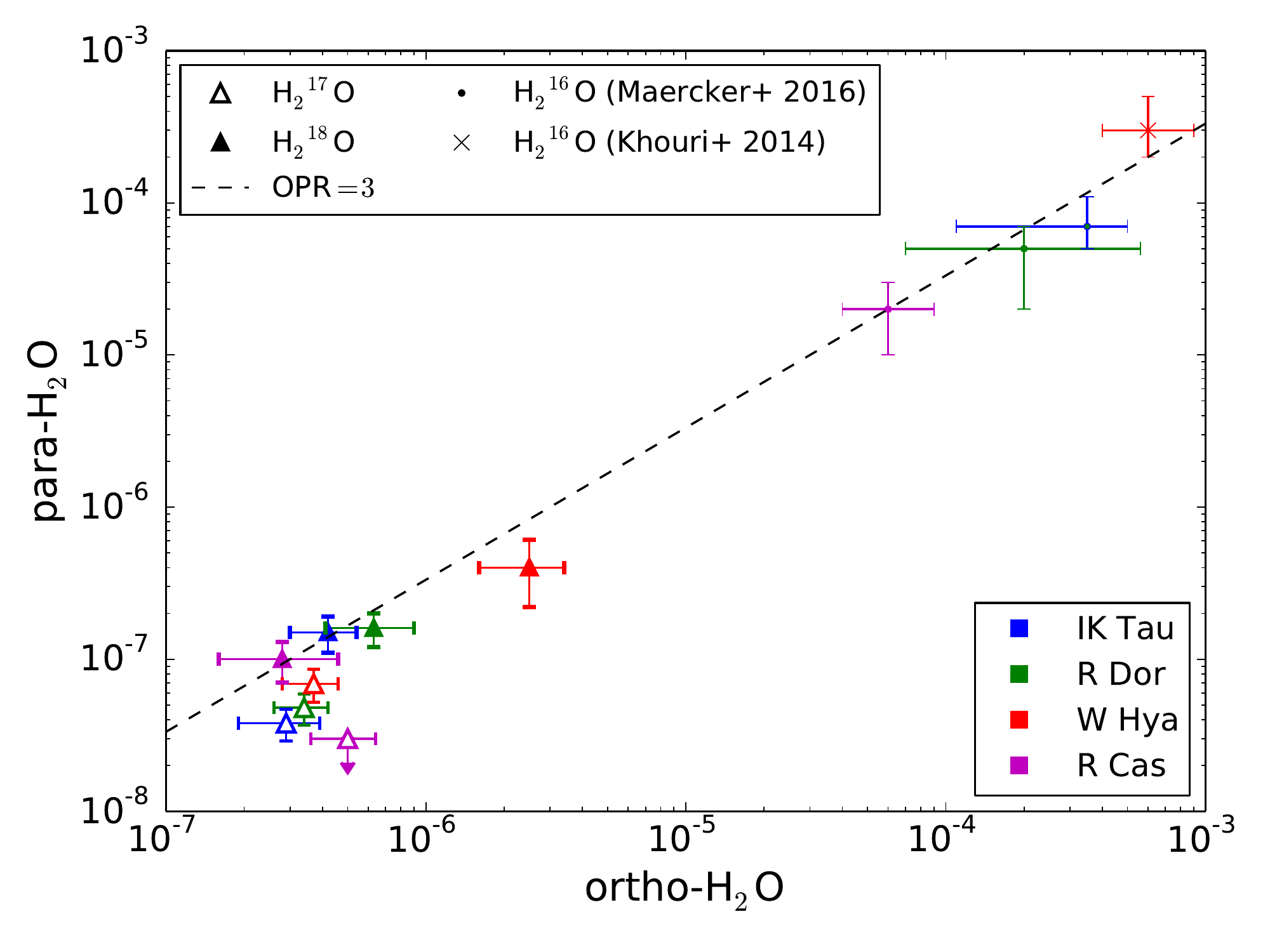}
\caption{{\it Left:} A visual representation of the \h2O ortho-to-para ratios, separated into \h2\up{17}O (open triangles) and \h2\up{18}O (filled triangles) isotopologues and colour-coded by source (see lower legend). The dotted line indicates the expected ortho-to-para ratio of 3 (see text for more details). {\it Right:} A plot of the same data as shown in the left panel, with the addition of \h2\up{16}O data taken from \cite{Maercker2016} and \cite{Khouri2014a}.}
\label{OPRplot}
\end{center}
\end{figure*}

A visual representation of the ortho-to-para ratios (OPR) for all modelled stars and isotopologues is plotted in Fig. \ref{OPRplot}. As can be seen there, most of the datapoints fall close to the expected OPR of 3, within error margins. The most significant outlier is R~Cas \h2\up{17}O, for the reasons discussed above. In general, the \h2\up{18}O results are more consistent with the expected OPR of 3, possibly because the stronger \h2\up{18}O lines allow for more accurately determined models. Similar OPR results close to 3 were found by previous studies for \h2\up{16}O in the same sample, shown in the right panel of Fig. \ref{OPRplot}. The numerical values of the OPRs are listed in the last two lines in Table \ref{results}.

\section{Discussion}

\subsection{Comparison with other studies}

Two stars in our sample have been previously modelled using a different radiative transfer code, GASTRoNOoM: IK~Tau by \cite{Decin2010a} and W~Hya by \cite{Khouri2014a}. In both studies, the \h2\up{17}O and \h2\up{18}O abundances were determined based solely on HIFI observations, without the PACS lines included in this study, meaning that the radiative transfer models for each isotopologue and spin isomer were only constrained by one observed line each. Nevertheless, the \h2\up{17}O/\h2\up{18}O ratio that \cite{Decin2010a} found for IK~Tau is 0.33, in agreement with our results, despite the absolute abundances for the various \h2O isotopologues and spin isomers differing significantly from our results and those of \cite{Maercker2016}. The difference in absolute abundances is probably due to the different mass-loss rates used in the two studies, with \cite{Decin2010a} using $8\e{-6}\spy$ compared with our value of $5\e{-6}\spy$. The difference in photodissociation radii, with \cite{Decin2010a} using a value more than twice that of our $e$-folding radius, would also have contributed to the difference in absolute abundances.
The absolute abundances found by \cite{Khouri2014a} for W~Hya also differ significantly from our models, with differences ranging from factors of a few to close to an order of magnitude in the case of o-\h2\up{18}O. They too find a p-\h2\up{17}O/p-\h2\up{18}O ratio in very good agreement with our result, although their o-\h2\up{17}O/o-\h2\up{18}O is an order of magnitude smaller. The difference in absolute abundances probably arises from some of the different assumptions in our two models. For example, although the velocity and abundance profiles used in the two studies are the same, the dust properties used differ slightly, resulting in different dust temperature profiles and hence radiation fields in the models.
As \cite{Khouri2014a} did, we also find the o-\h2\up{18}O ($3_{1,2}\to3_{0,3}$) to be shifted slightly bluewards in frequency, but we are able to otherwise reproduce the line shape and strength reasonably well with a model that is also in agreement with the observed PACS line and has an abundance about an order of magnitude lower than used by \cite{Khouri2014a}. 
Both of these comparisons show that although the two radiative transfer codes GASTRoNOoM and ALI might give different results for \h2O modelling in terms of absolute abundances, they generally give consistent results for isotopologue ratios modelled using consistent methods.

\cite{Hinkle2016} investigated the oxygen isotopic ratios for a large sample of AGB stars, using ro-vibrational CO lines in the 1.5--2.5 $\mic$ region and a curve of growth analysis method. From the stars in our sample, they determined \up{16}O/\up{17}O and \up{16}O/\up{18}O ratios for W~Hya and IK~Tau. Their ratios determined with respect to \up{16}O do not match our equivalent results, with a factor of a few differences. Converting these to \up{17}O/\up{18}O ratios, their study agrees with our result for W~Hya, but is almost a factor of 4 smaller than our result for IK~Tau.

\begin{figure*}[t]
\begin{center}
\includegraphics[width=0.49\textwidth]{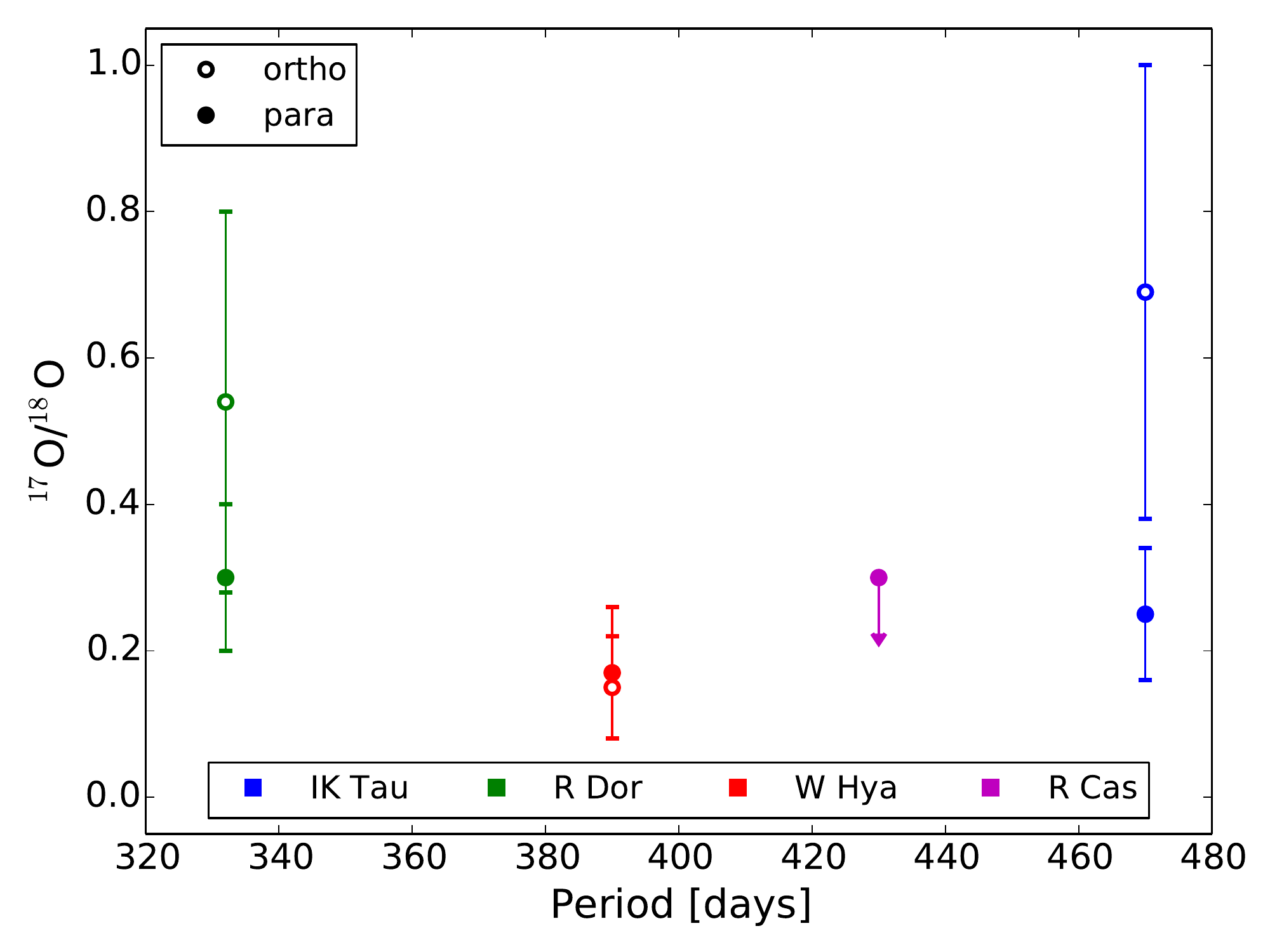}
\includegraphics[width=0.49\textwidth]{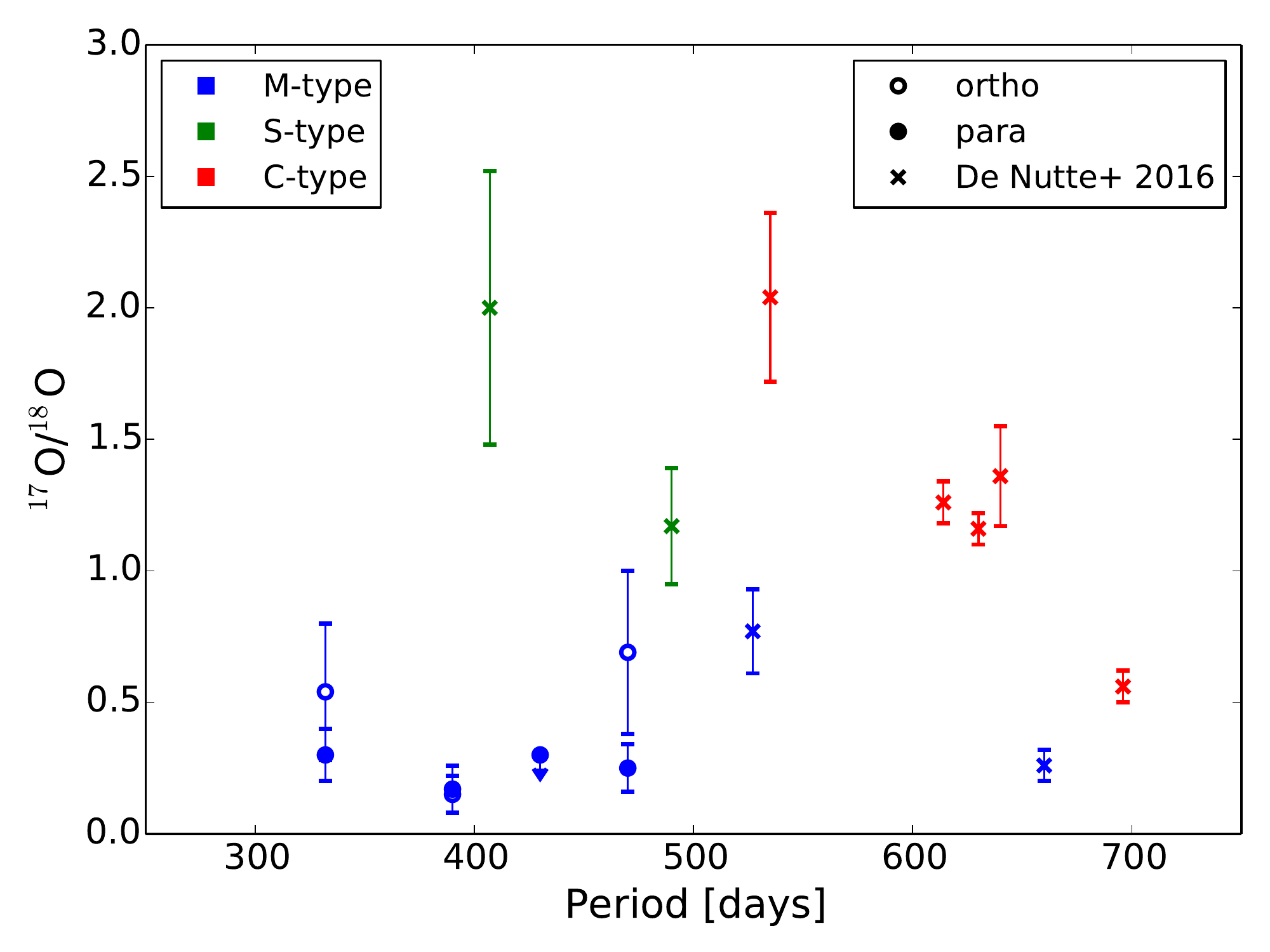}
\caption{{\it Left:} The \up{17}O/\up{18}O ratios plotted against the (primary) pulsational periods of our four sources {\it Right:} A plot of the same quantities but including the results of \cite{De-Nutte2016} and coloured by chemical type.}
\label{ratiovsperiod}
\end{center}
\end{figure*}

\cite{De-Nutte2016} investigated the \up{17}O/\up{18}O ratios for a different sample of AGB stars that did not overlap with ours. Their result showed a tentative inverse trend in \up{17}O/\up{18}O ratio against period for a small sample of chemically diverse AGB stars. As can be seen in the left panel of Fig. \ref{ratiovsperiod}, there is no clear trend between \up{17}O/\up{18}O ratio and period for our sources. In the right panel of Fig. \ref{ratiovsperiod} we plot our results with those of \cite{De-Nutte2016}, differentiating between chemical types. There it can be seen that our results negate any apparent inverse trend with period, which was most likely a coincidental function of the chosen sources. The tendency for M-type (non-OH/IR) AGB stars to have generally lower \up{17}O/\up{18}O ratios than other chemical types is supported by our results. However, this trend is far from certain and a larger sample size is required to confirm it and to be able to draw any firm conclusions.

\subsection{Determination of initial stellar mass from \up{17}O/\up{18}O ratios}\label{massdet}

As discussed in \cite{Lattanzio2003}, \cite{Karakas2014}, \cite{De-Nutte2016}, and referenes therein, \up{17}O/\up{18}O ratios are linked to the initial masses of AGB stars that have not experienced hot bottom burning. This is because the surface abundances of these two isotopes are altered by first dredge up, which occurs during the RGB phase, to an extent dependent on the initial mass, but are not significantly changed during second or third dredge ups. Hence the \up{17}O/\up{18}O ratio is a marker of the star's initial mass.

Based on models from \cite{Stancliffe2004}, \cite{Karakas2016}, and \cite{Cristallo2011}, \cite{De-Nutte2016} compared their \up{17}O/\up{18}O ratios, derived from observations of C\up{17}O and C\up{18}O, with stellar evolution model predictions to determine the initial masses of a chemically diverse sample of AGB stars. There is some uncertainty in mass determinations for \up{17}O/\up{18}O $\gtrsim 1.5$, where the function is not injective, but our model results give ratios lower than this, making mass determinations more straightforward. In the regime we are interested in, metallicity does not appear to play a significant role, as can be seen in Fig. 4 of \cite{De-Nutte2016}. Taking our error margins into account and assuming that \h2\up{17}O/\h2\up{18}O ratios are directly representative of the \up{17}O/\up{18}O ratios, we estimate the following initial masses of our stars using Fig. 2 in \cite{De-Nutte2016}, with the results also listed in Table \ref{masses}. { For IK~Tau we find initial mass estimates in the range $1.1 \lesssim M_\mathrm{initial}\lesssim1.5~\msol$.} For R~Dor we find possible initial masses of $1.0 \lesssim M_\mathrm{initial}\lesssim1.3~\msol$. The upper limit for the \up{17}O/\up{18}O ratio derived from only the p-\h2O results for R~Cas (see Sect. \ref{sec:results}) gives an initial mass upper limit of $\lesssim 1.1~\msol$. We find lower \h2\up{17}O/\h2\up{18}O ratios for W~Hya than covered by the evolutionary models, suggesting a very low initial mass, $0.8 \lesssim M_\mathrm{initial}\lesssim1~\msol$. However, the H-burning lifetime of a $0.8~\msol$ star with solar metallicity, $Z_\odot=0.014$, is 24~Gyr and for low metallicity, $Z=0.004$, is 20~Gyr, ruling out such a low initial mass for W~Hya except at very low metallicities. We also note that 0.19 is the solar \up{17}O/\up{18}O ratio \citep{Asplund2009} and is used as the initial ratio by the \cite{Karakas2016} models for all metallicities, hence leading to higher ratios after the first dredge up.

\begin{table}[tp]
\caption{Initial mass estimates.}\label{masses}
\begin{center}
\begin{tabular}{cc}
\hline\hline
Star & Initial mass [$\msol$]\\
\hline
IK~Tau & $1.1 \lesssim M_\mathrm{initial}\lesssim1.5$\\
R~Dor & $1.0 \lesssim M_\mathrm{initial}\lesssim1.3$\\
W~Hya & $\approx1.0$\\
R~Cas & $\lesssim 1.1$\\
\hline																				
\end{tabular}
\tablefoot{ Derived from results presented in \cite{De-Nutte2016}. See text for details.}
\end{center}
\end{table}%

There is no concrete reason to assume that the solar \up{16}O/\up{17}O and \up{16}O/\up{18}O ratios, upon which the \up{17}O/\up{18}O ratio is based, should apply to all main sequence stars. Studies of the oxygen isotopic ratios across the galaxy have shown that the solar system is an outlier when compared with various molecular clouds and star-forming regions. For example, \cite{Penzias1981} found a consistent value of \up{17}O/\up{18}O~$\approx0.3$ across giant molecular clouds in different parts of the galaxy. The only two outliers in the study were the stellar sources: the solar system (lower than average) and the carbon star CW Leo (higher than average). 
%The average \up{17}O/\up{18}O results found by \cite{Penzias1981} and by the more recent study of \cite{Wouterloot2008} match the average of our small M-type sample, but of course . 
Gradients in \up{17}O/\up{18}O ratios with galactic radius have been found by \cite{Wouterloot2008} and others, indicating higher ratios close to the galactic centre and lower ratios in the outer disc. This may not be a function of metallicity, however, since the metal-poor Large Magellanic Cloud has larger average values (\up{17}O/\up{18}O~$\approx0.7$) than molecular clouds in the Milky Way, including those in the outer disc where metallicities are comparable to the LMC \citep{Heikkila1998}. In any case, since evolved stars contribute to the chemical enrichment of the galaxy, there is no reason to assume their \up{17}O/\up{18}O ratios should correlate with those found for present-day molecular clouds, since these might not have significant bearing on the formation conditions of the main sequence progenitors of AGB stars.

Modifying the initial \up{17}O/\up{18}O ratio for the $1~\msol$ and solar metallicity model from \cite{Karakas2016} also shifts the \up{17}O/\up{18}O ratio found after the first dredge up and at the first thermal pulse. The results of these tests are shown in Table \ref{ratiotests} and highlight the uncertainty introduced into the initial mass estimate by the choice of initial \up{17}O/\up{18}O ratio. Note also the slight increase in \up{17}O/\up{18}O ratio at the first thermal pulse, which is actually due to some \up{17}O (and \up{13}C) being dredged up during the early AGB when the convective envelope moves inwards. Subsequent thermal pulses have only a small impact on the \up{17}O/\up{18}O ratio, especially for low \up{17}O/\up{18}O ratios and low initial masses, as is shown in Figures 5--7 of \cite{Karakas2016}. Hence, W~Hya, with its low \up{17}O/\up{18}O could have had an initial mass of $1~\msol$ or even slightly higher, depending on the initial \up{17}O/\up{18}O ratio it had when it entered the main sequence. Similar uncertainties apply to the other stars in our sample. 
 Nevertheless, in the absence of clear constraints on initial \up{17}O/\up{18}O ratios, the initial masses listed in Table \ref{masses} serve as good approximations for our sample of low-mass M-type AGB stars.

\begin{table}[tp]
\caption{Variations in \up{17}O/\up{18}O ratios after first dredge-up and immediately prior to the first thermal pulse for differing initial ratios.}\label{ratiotests}
\begin{center}
\begin{tabular}{c|ccc}
\hline\hline
Initial mass  &Initial  & Post-FDU &     First TP\\
$[\msol]$ \\
\hline
1.0 &     0.100          &        0.108 & 0.114\\
1.0 &     0.150          &      0.164 &            0.172\\
1.0 &     \phantom{*}0.190*   &  0.207   &          0.291\\
0.8 & \phantom{*}0.190*   &0.191 & 0.213\\
\hline																				
\end{tabular}
\tablefoot{* is the solar ratio. { Model results are from \cite{Karakas2016}.}}
\end{center}
\end{table}%

\section{Conclusions}

We performed detailed radiative transfer models of \h2\up{17}O and \h2\up{18}O for a sample of four M-type AGB stars. These models, constrained by \textsl{Herschel}/HIFI and \textsl{Herschel}/PACS observations, indicate that the observed lines are all optically thick, despite the relative rarity of the studied isotopologues, meaning that radiative transfer modelling, rather than a comparison of line intensities, is the only reliable way to determine abundances and abundance ratios. For o-\h2\up{18}O towards IK~Tau we had a large number of available lines to constrain our models, but for the other sources we were generally limited to one or two lines per isotopologue and spin isomer, unfortunately leading to less precise models.

Overall, we found lower abundances of \h2\up{17}O than \h2\up{18}O, indicating that the stars in our sample have not undergone hot bottom burning, as was expected given that they have not otherwise been identified as OH/IR stars. We found rather low \h2\up{17}O/\h2\up{18}O ratios which, assuming a direct conversion to \up{17}O/\up{18}O ratios, indicate that all the stars in our sample had relatively low initial masses, in the range $\sim1.0$ to 1.5 $\msol$.

The ortho to para ratios we found for the two studied isotopologues were close to the expected value of 3, but occasionally a bit higher, probably due to the low numbers of observed lines available to constrain most of our models.

\begin{acknowledgements}

LD, TD acknowledge support from the ERC consolidator grant 646758 AEROSOL and the FWO Research Project grant G024112N. HO acknowledges financial support from the Swedish Research Council.

HIFI has been designed and built by a consortium of institutes and university departments from across Europe, Canada and the United States under the leadership of SRON Netherlands Institute for Space Research, Groningen, The Netherlands and with major contributions from Germany, France and the US. Consortium members are: Canada: CSA, U.Waterloo; France: CESR, LAB, LERMA, IRAM; Germany: KOSMA, MPIfR, MPS; Ireland, NUI Maynooth; Italy: ASI, IFSI-INAF, Osservatorio Astrofisico di Arcetri-INAF; Netherlands: SRON, TUD; Poland: CAMK, CBK; Spain: Observatorio Astron\'omico Nacional (IGN), Centro de Astrobiolog\'ia (CSIC-INTA). Sweden: Chalmers University of Technology - MC2, RSS \& GARD; Onsala Space Observatory; Swedish National Space Board, Stockholm University - Stockholm Observatory; Switzerland: ETH Zurich, FHNW; USA: Caltech, JPL, NHSC.

PACS has been developed by a consortium of institutes led by MPE (Germany) and including UVIE (Austria); KU Leuven, CSL, IMEC (Belgium); CEA, LAM (France); MPIA (Germany); INAF-IFSI/OAA/OAP/OAT, LENS, SISSA (Italy); IAC (Spain). This development has been supported by the funding agencies BMVIT (Austria), ESA-PRODEX (Belgium), CEA/CNES (France), DLR (Germany), ASI/INAF (Italy), and CICYT/MCYT (Spain).

\end{acknowledgements}

%\begin{thebibliography}{}
%
%\end{thebibliography}

\bibliographystyle{aa}
\bibliography{MstarH2Oiso}

\Online
\appendix
\section{Plots of results}

\begin{figure}[t]
\begin{center}
\includegraphics[width=0.5\textwidth]{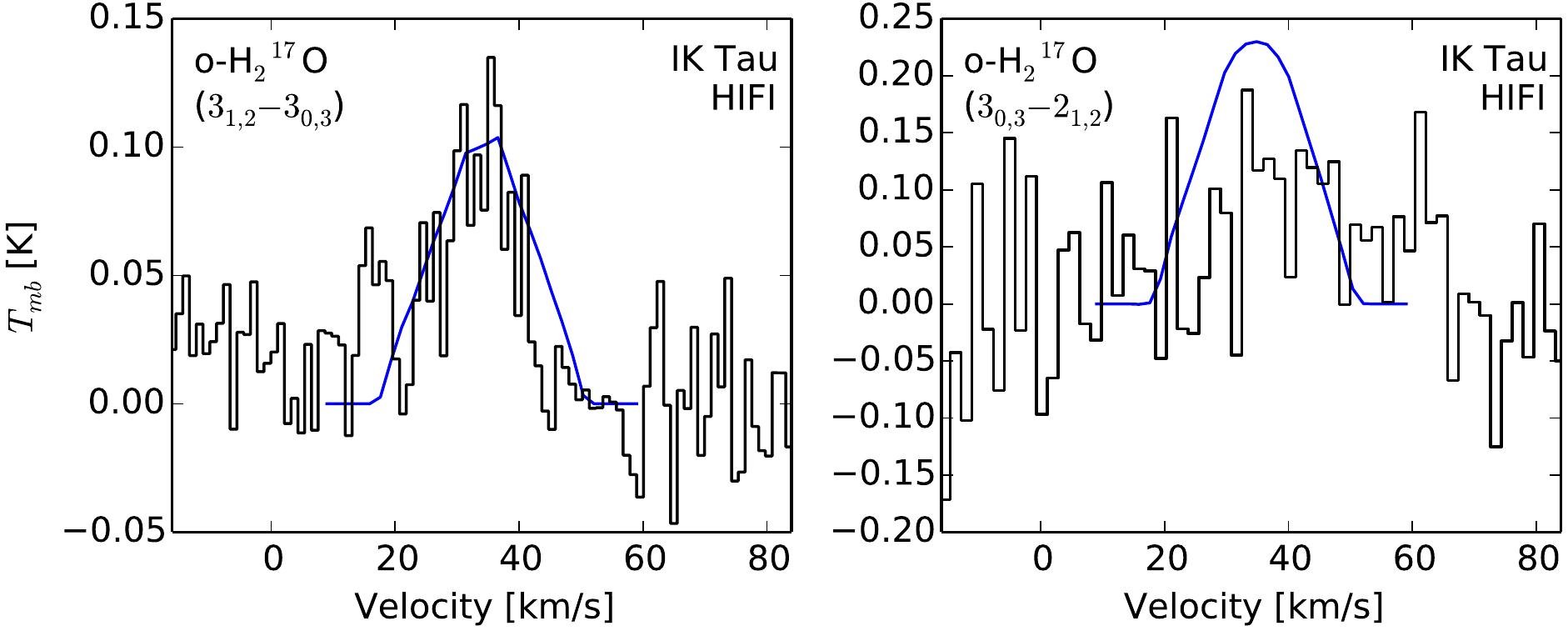}
\includegraphics[width=0.25\textwidth]{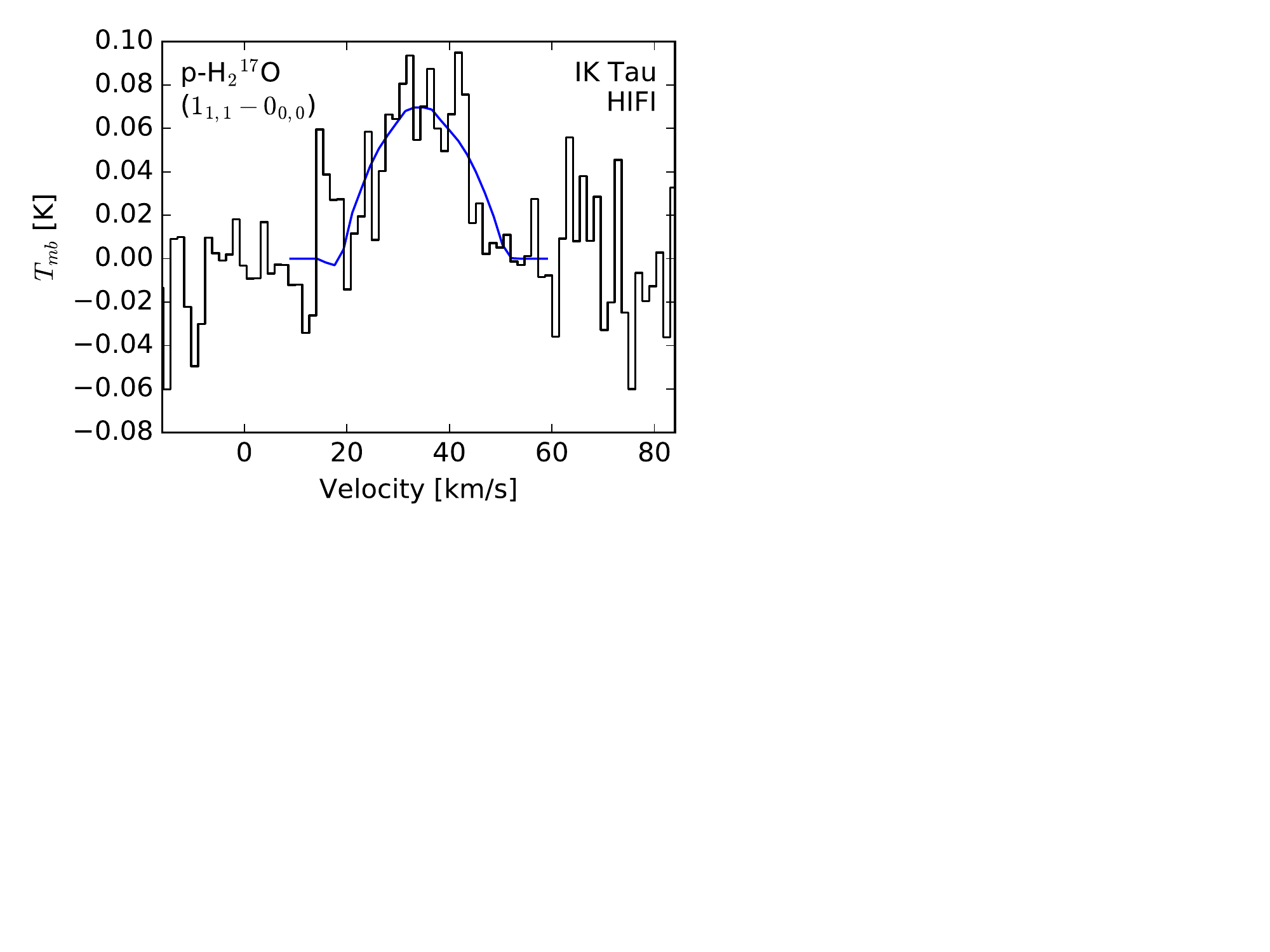}
\includegraphics[width=0.242\textwidth]{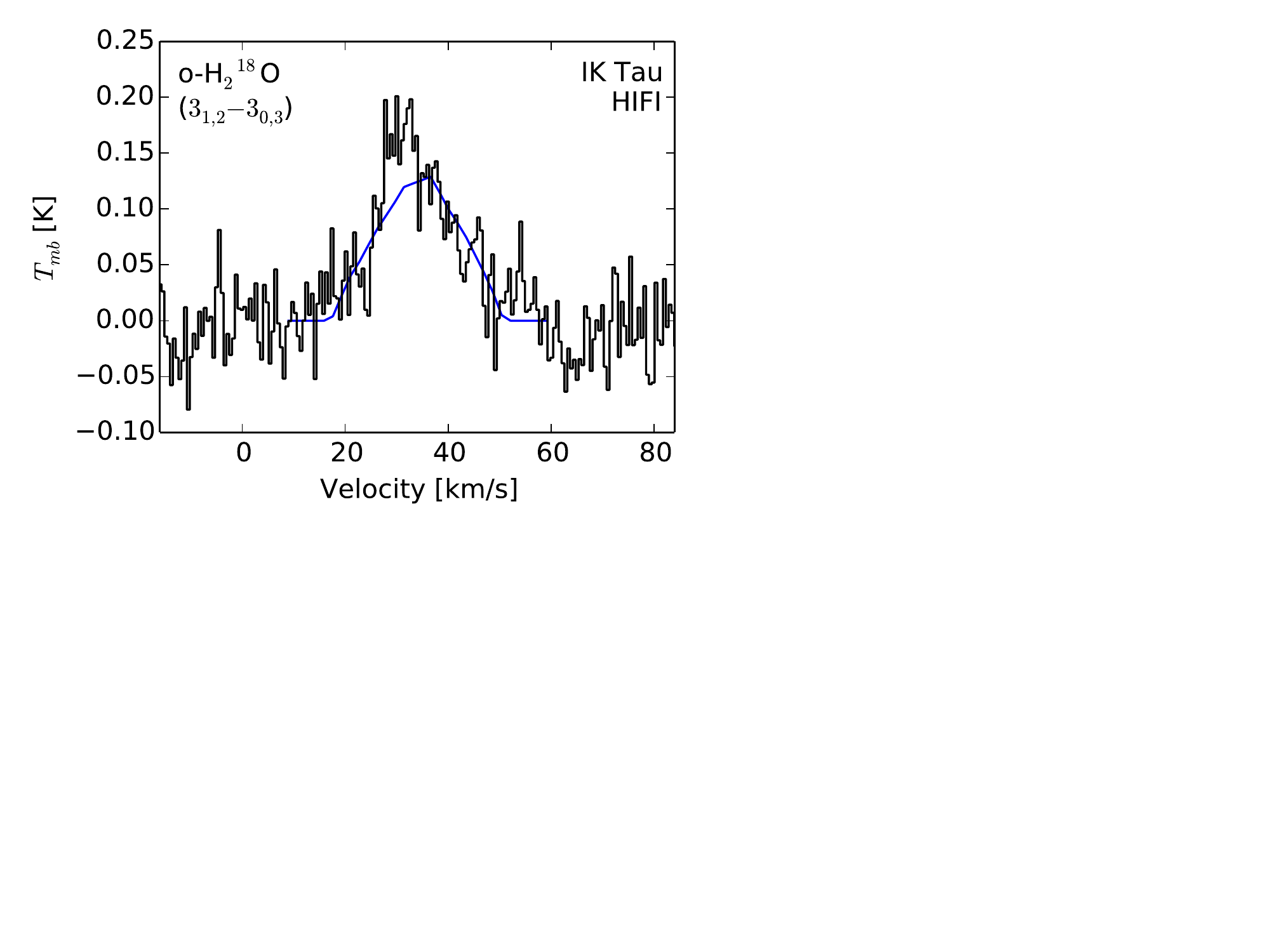}
\includegraphics[width=0.242\textwidth]{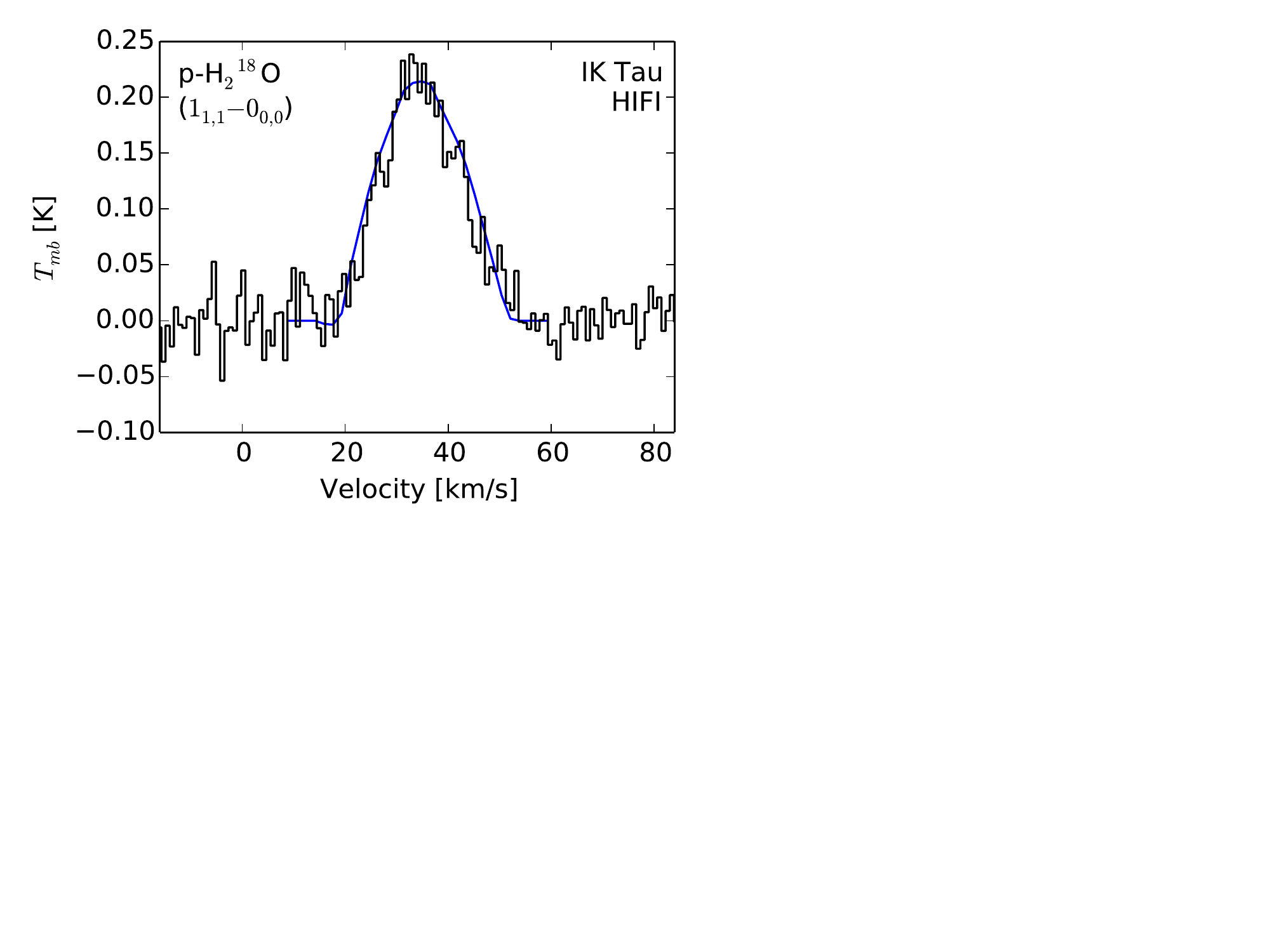}
\caption{HIFI lines (black histograms) and models (blue curves) for IK Tau.}
\label{iktauplots}
\end{center}
\end{figure}

Figures \ref{iktauplots}, \ref{rdorplots}, \ref{whyaplots}, and \ref{rcasplots} show the detected HIFI lines along with the corresponding model lines for IK~Tau, R~Dor, W~Hya, and R~Cas, respectively. For IK~Tau and R~Cas we have also included the section of spectrum showing the non-detected lines that we used to further constrain our models, along with the model line for those transitions.

Fig. \ref{fitplots} contains goodness of fit plots, showing the ratio between modelled integrated intensity (for HIFI) or flux (for PACS) and the observed quantity, plotted against the energy of the upper level of the transition. As can be seen, there are no clear trends with energy across these plots, although there is some scatter resulting from models that don't fit all the observed lines equally well.

\begin{figure}[t]
\begin{center}
\includegraphics[width=0.5\textwidth]{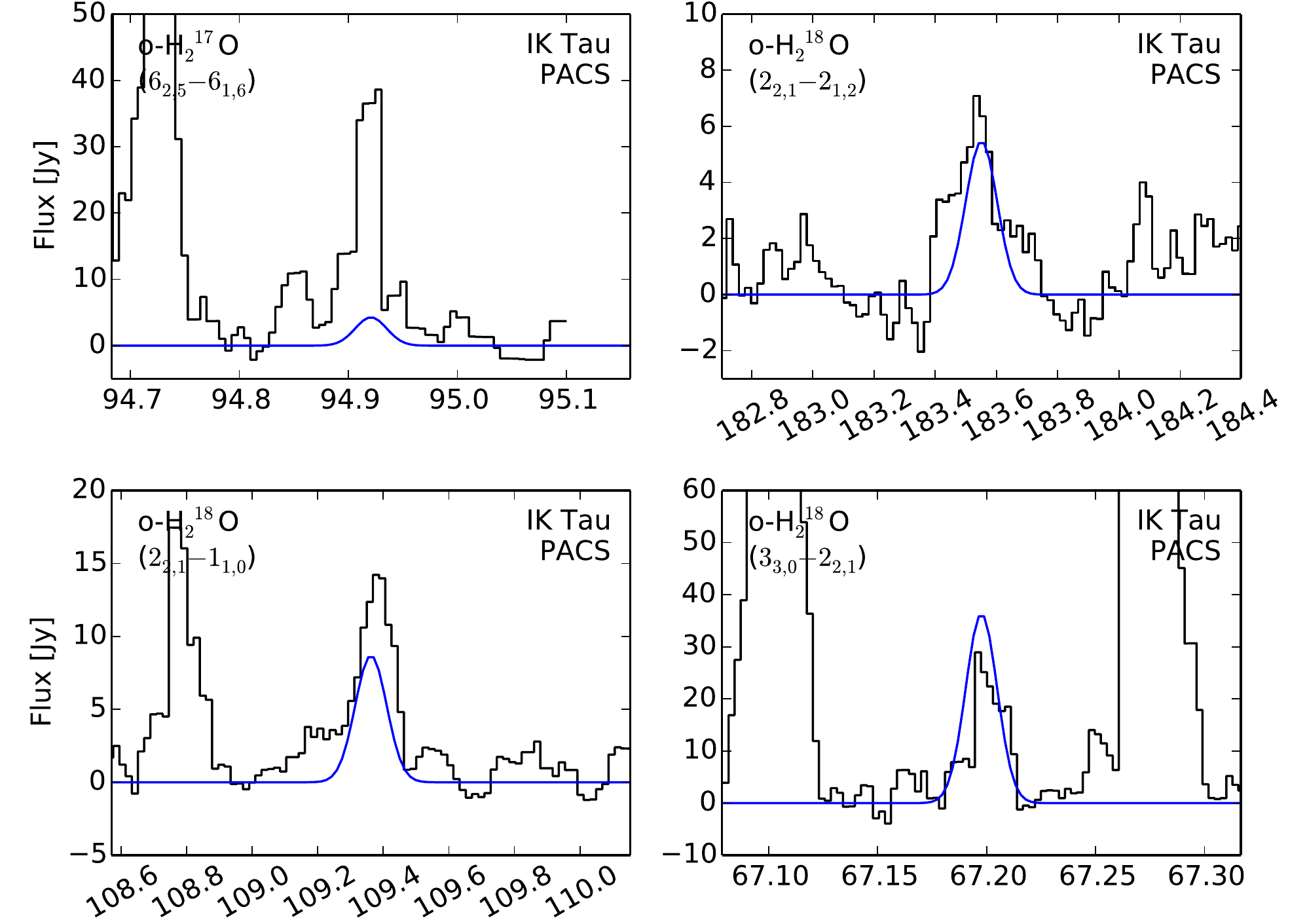}
\includegraphics[width=0.5\textwidth]{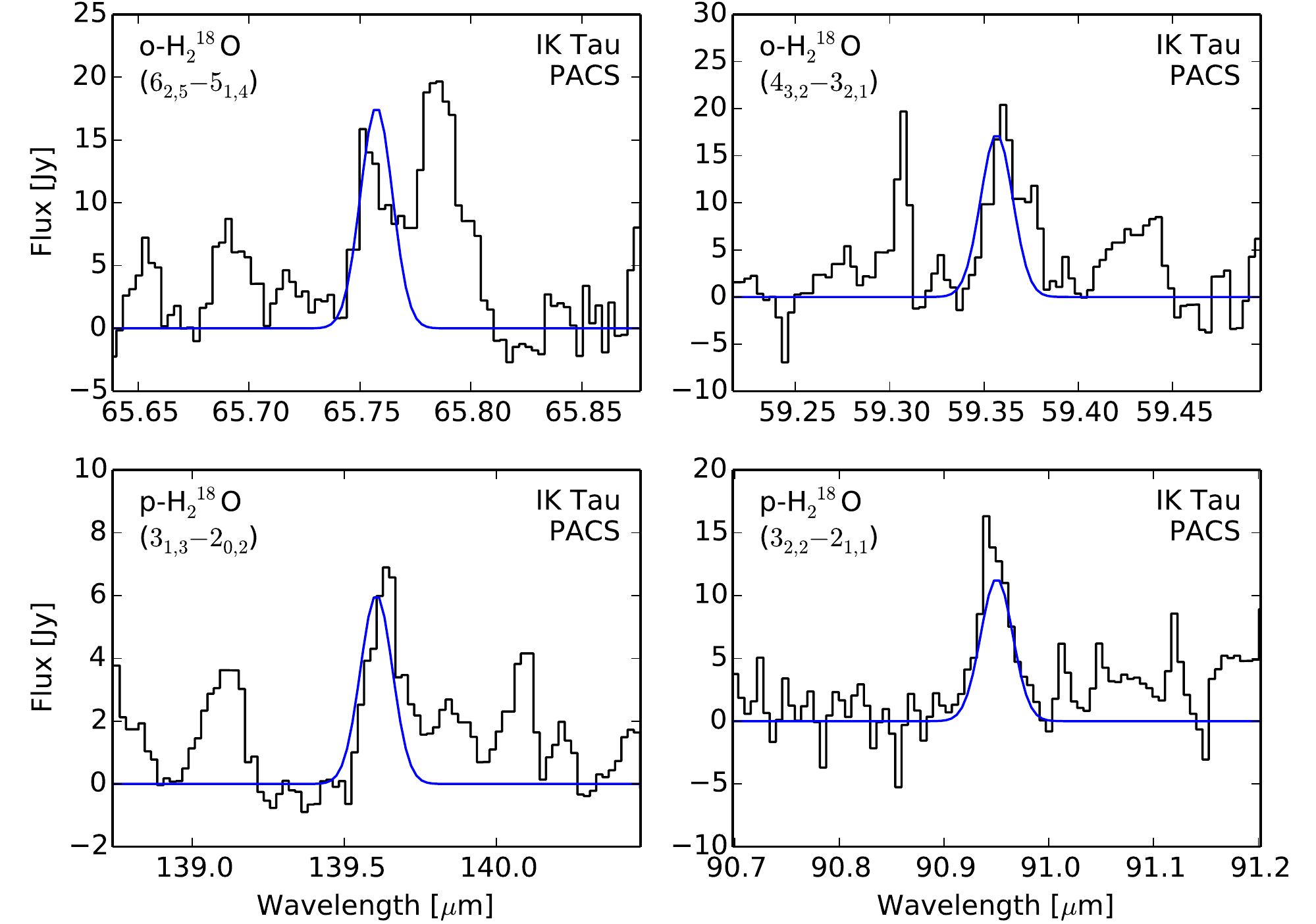}
\caption{ PACS lines (black histograms) and models (blue curves) for IK Tau.}
\label{iktaupacs}
\end{center}
\end{figure}

\begin{figure}[t]
\begin{center}
\includegraphics[width=0.242\textwidth]{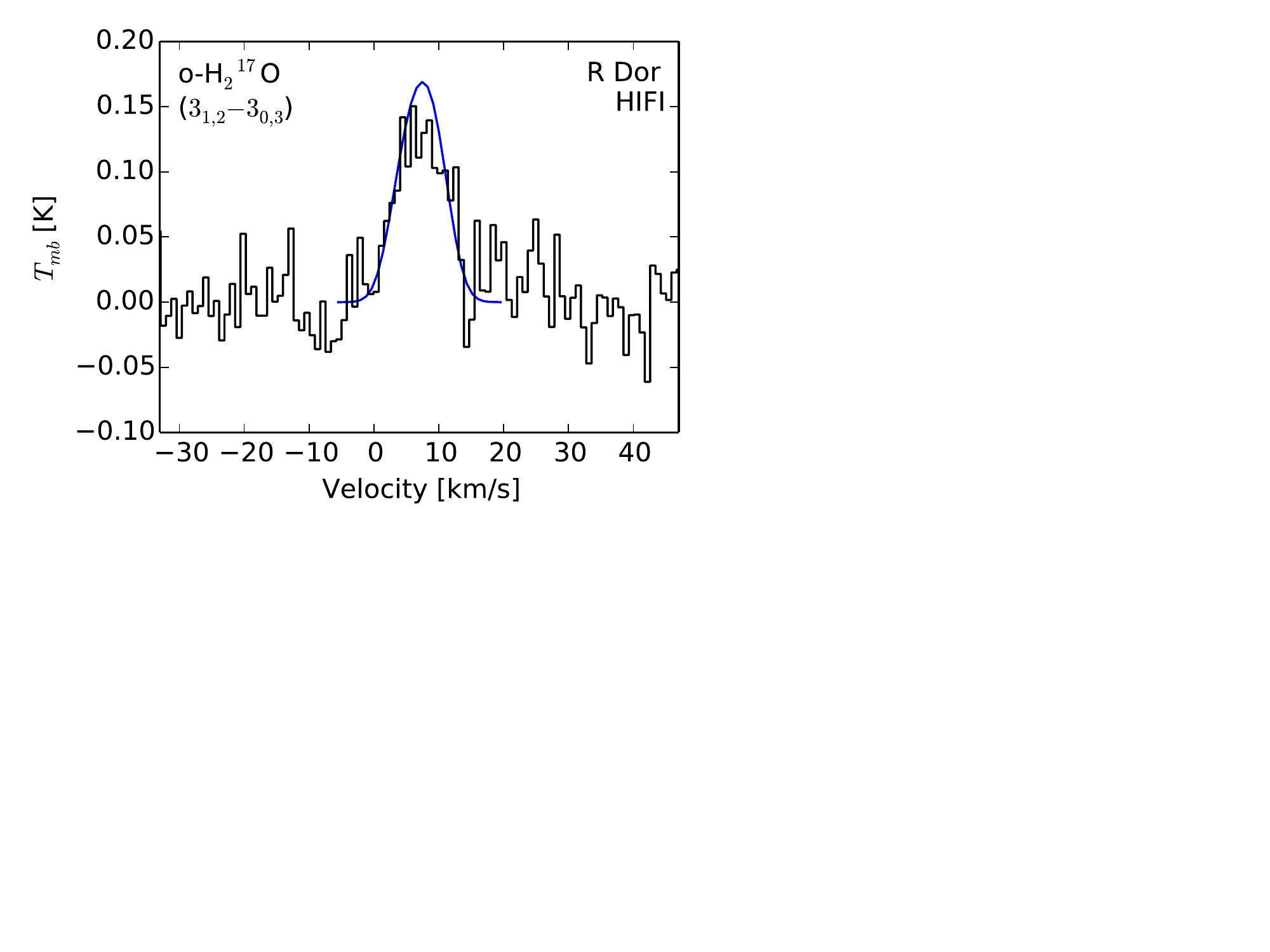}
\includegraphics[width=0.242\textwidth]{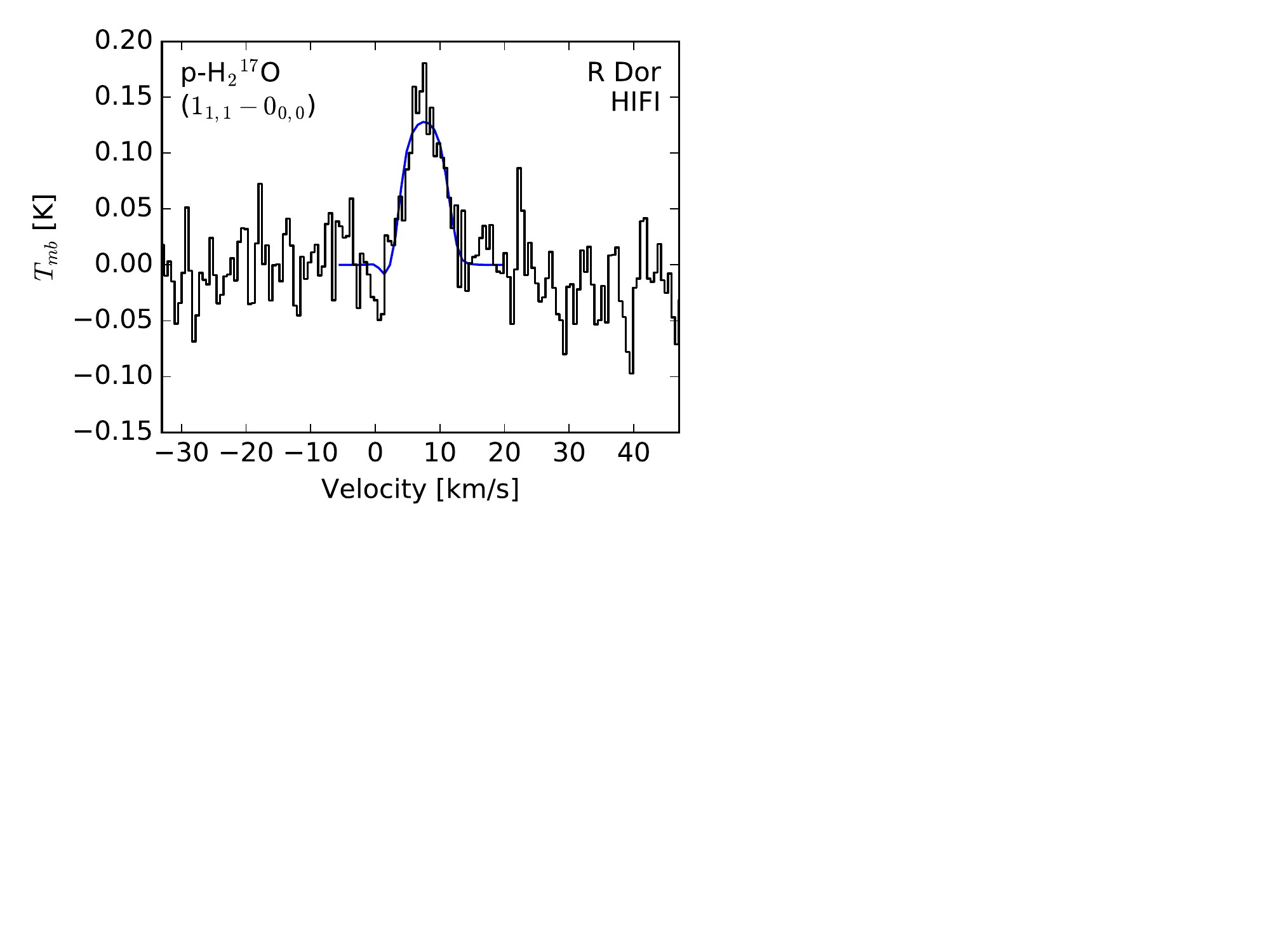}
\includegraphics[width=0.242\textwidth]{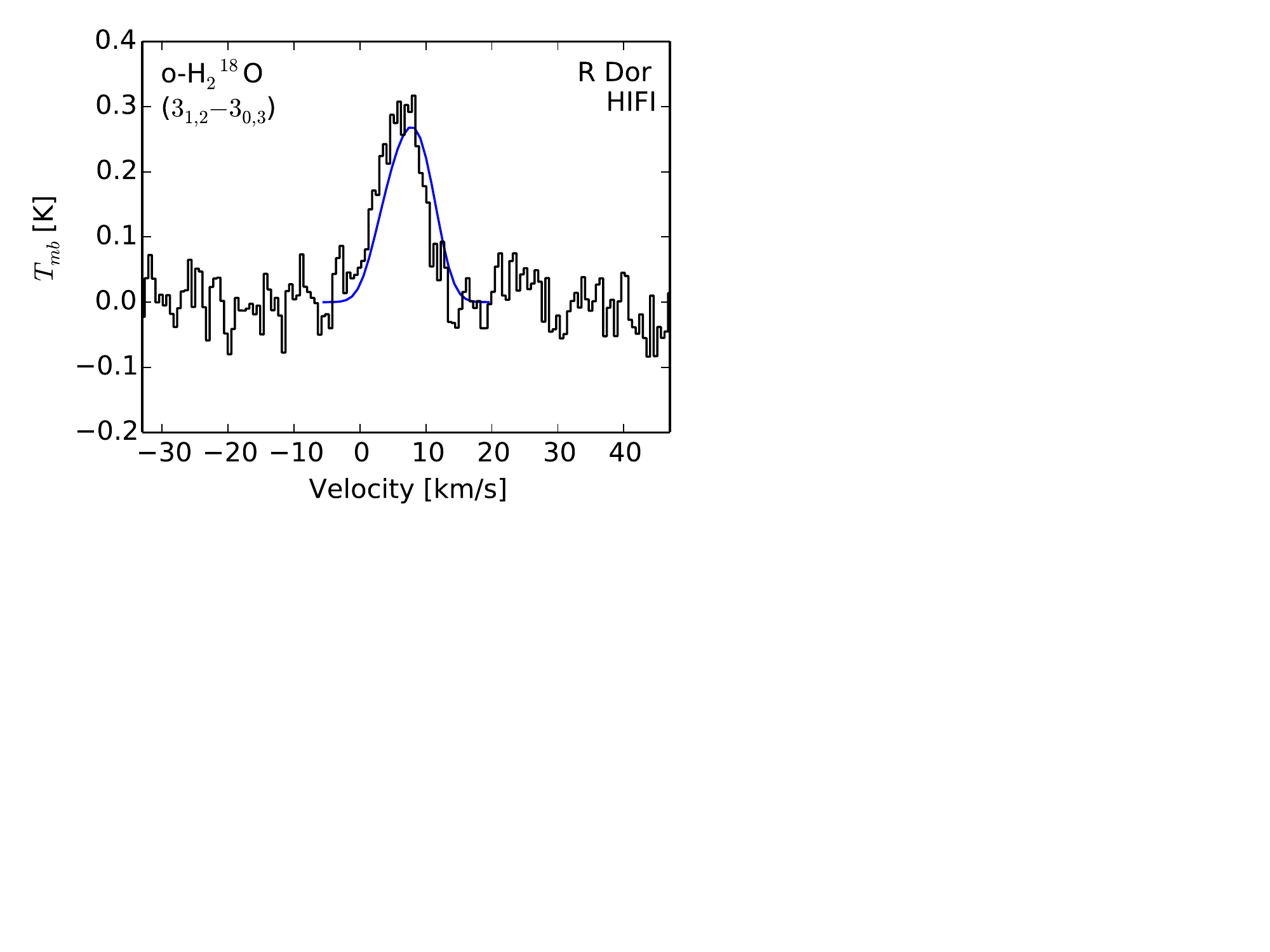}
\includegraphics[width=0.242\textwidth]{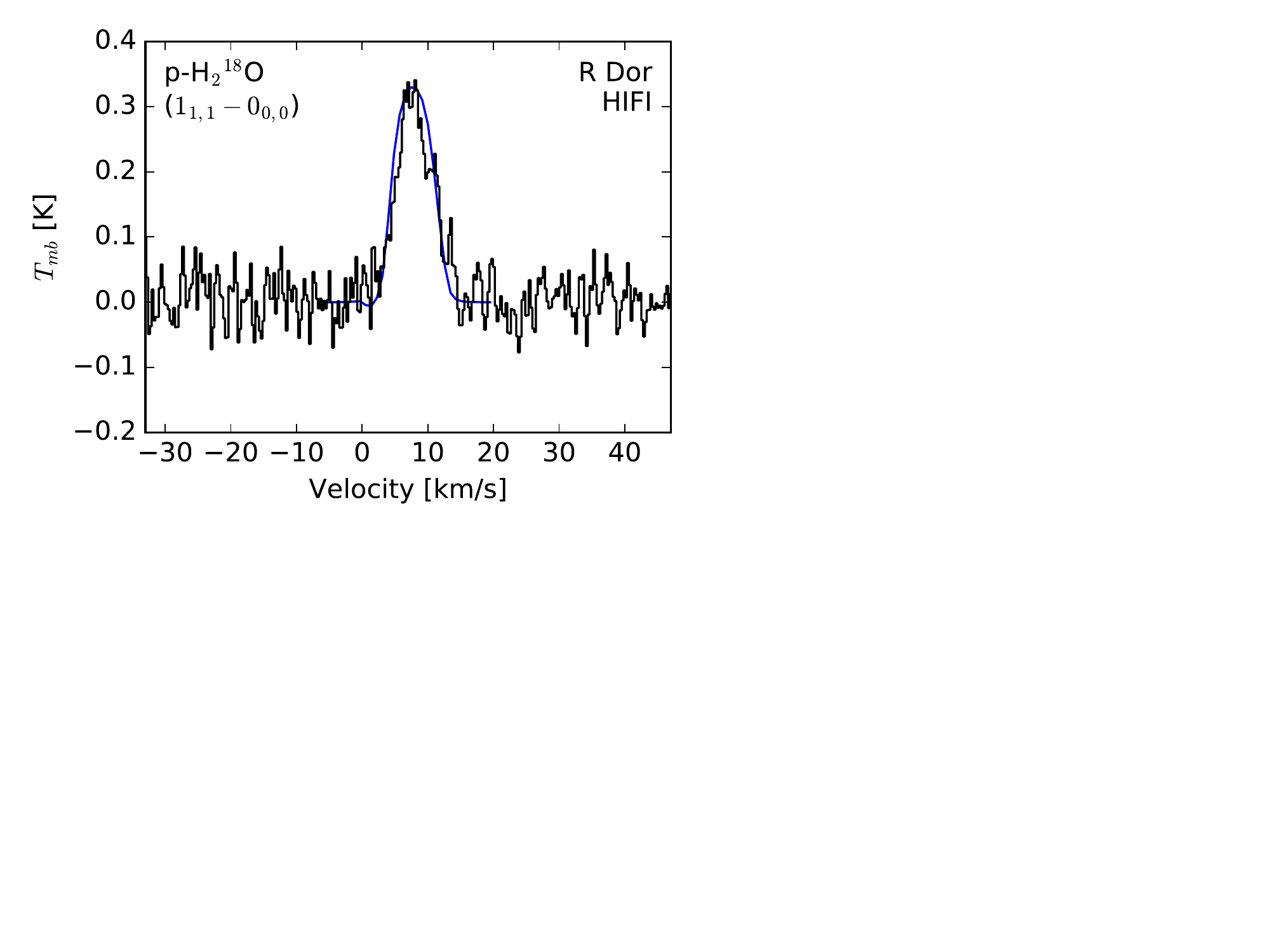}
\caption{HIFI lines (black histograms) and models (blue curves) for R Dor.}
\label{rdorplots}
\end{center}
\end{figure}
\begin{figure}[t]
\begin{center}
\includegraphics[width=0.25\textwidth]{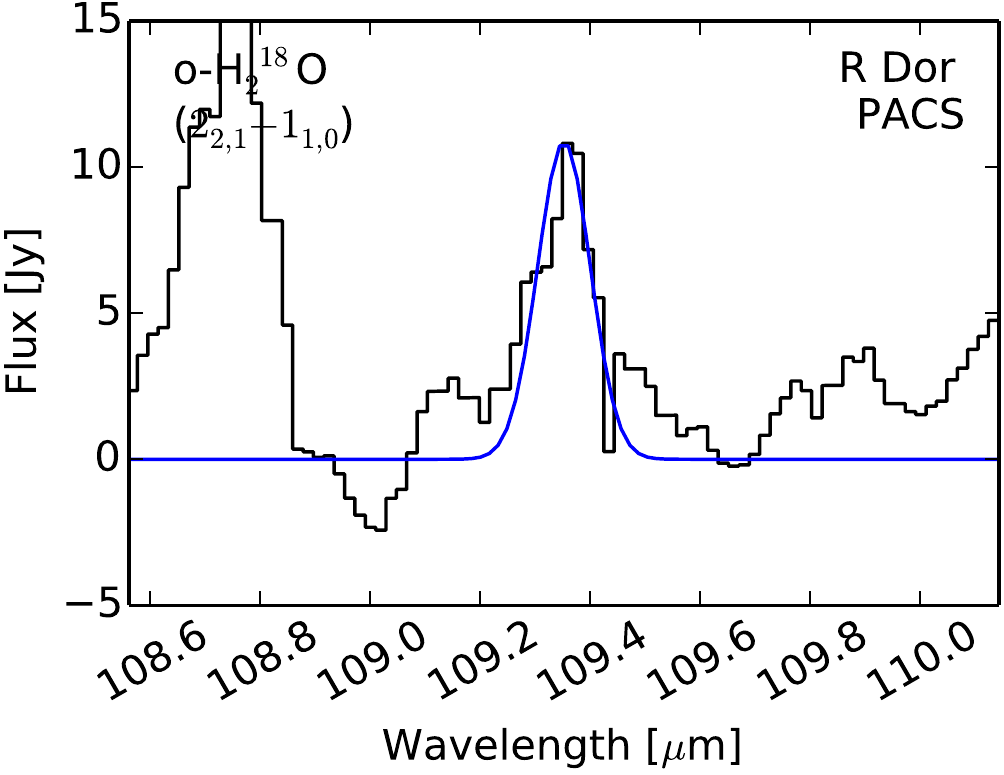}
\caption{ PACS line (black histogram) and model (blue curve) for R Dor.}
\label{rdorpacs}
\end{center}
\end{figure}

\begin{figure}[t]
\begin{center}
\includegraphics[width=0.5\textwidth]{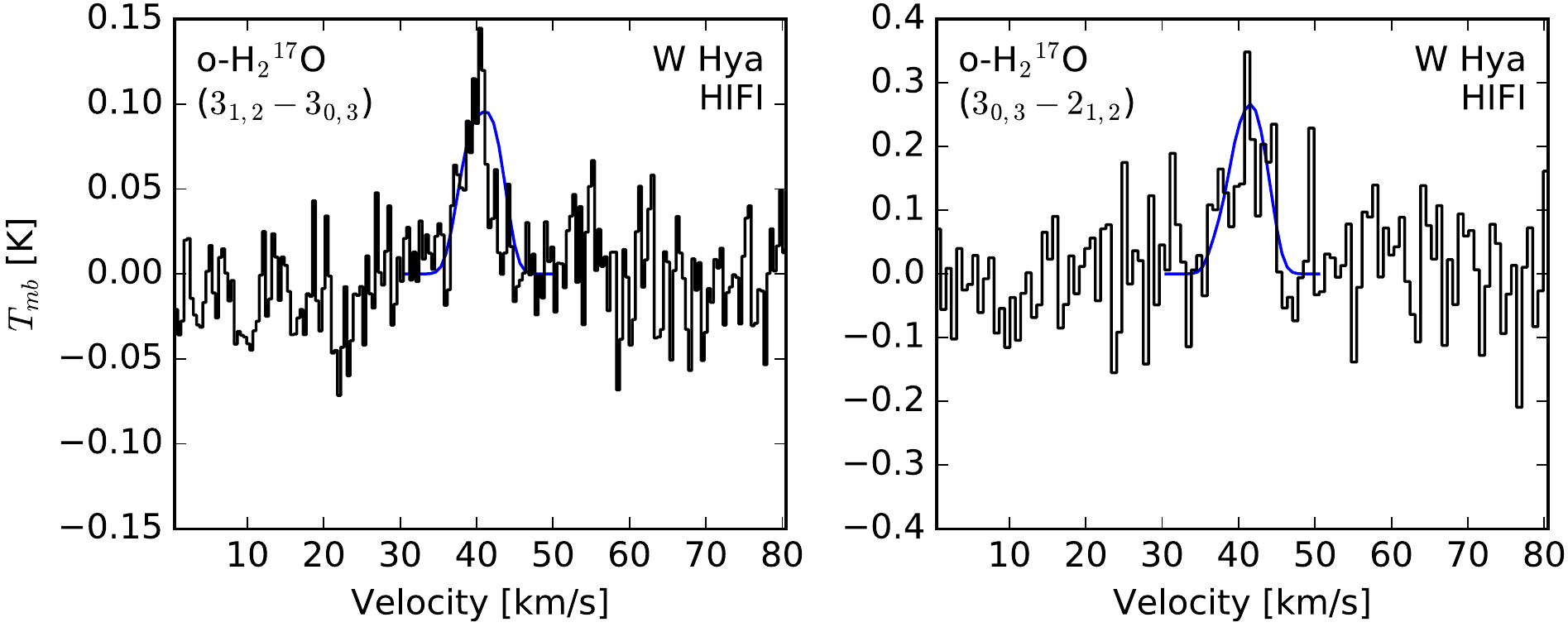}
\includegraphics[width=0.25\textwidth]{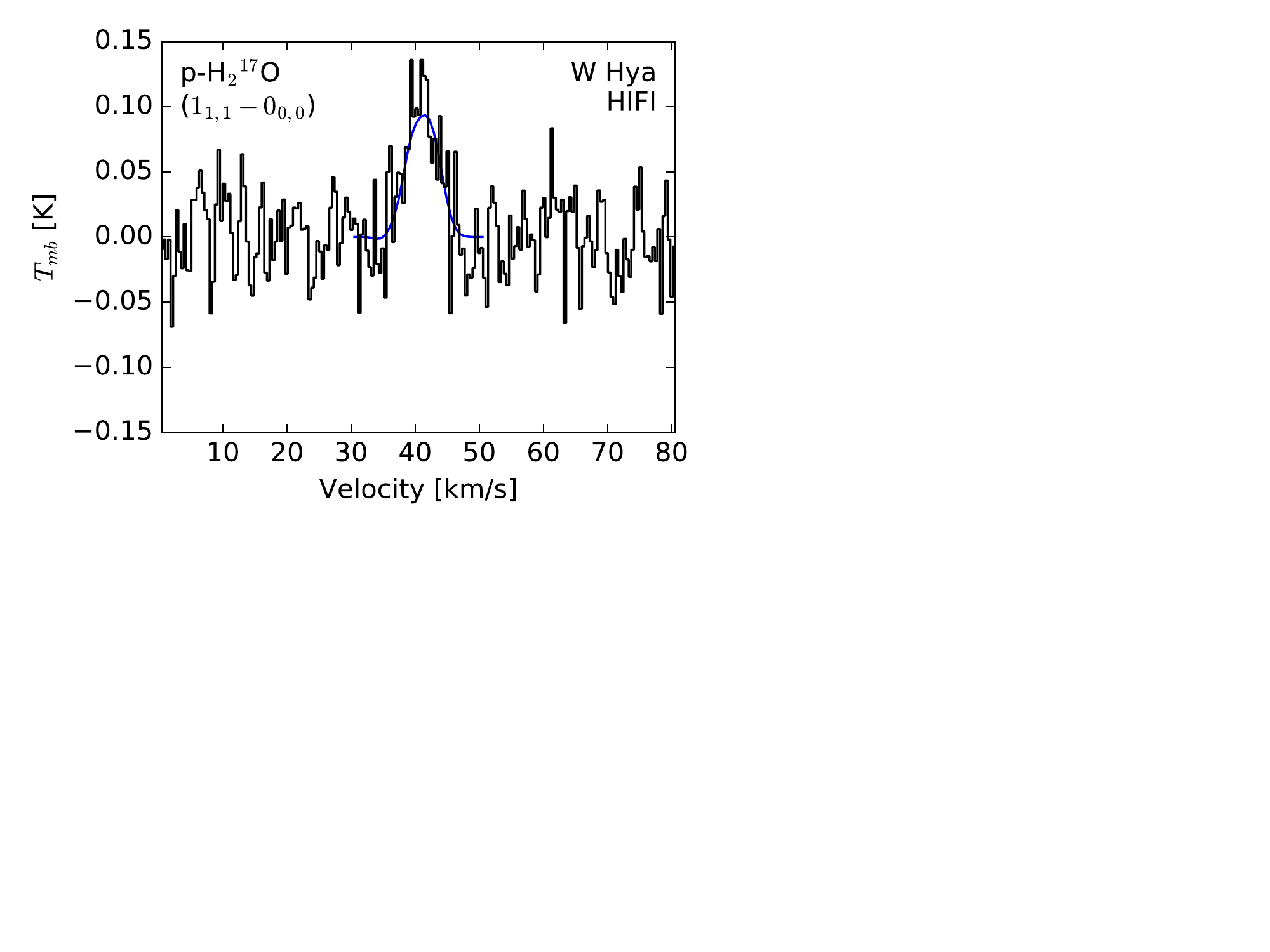}
\includegraphics[width=0.242\textwidth]{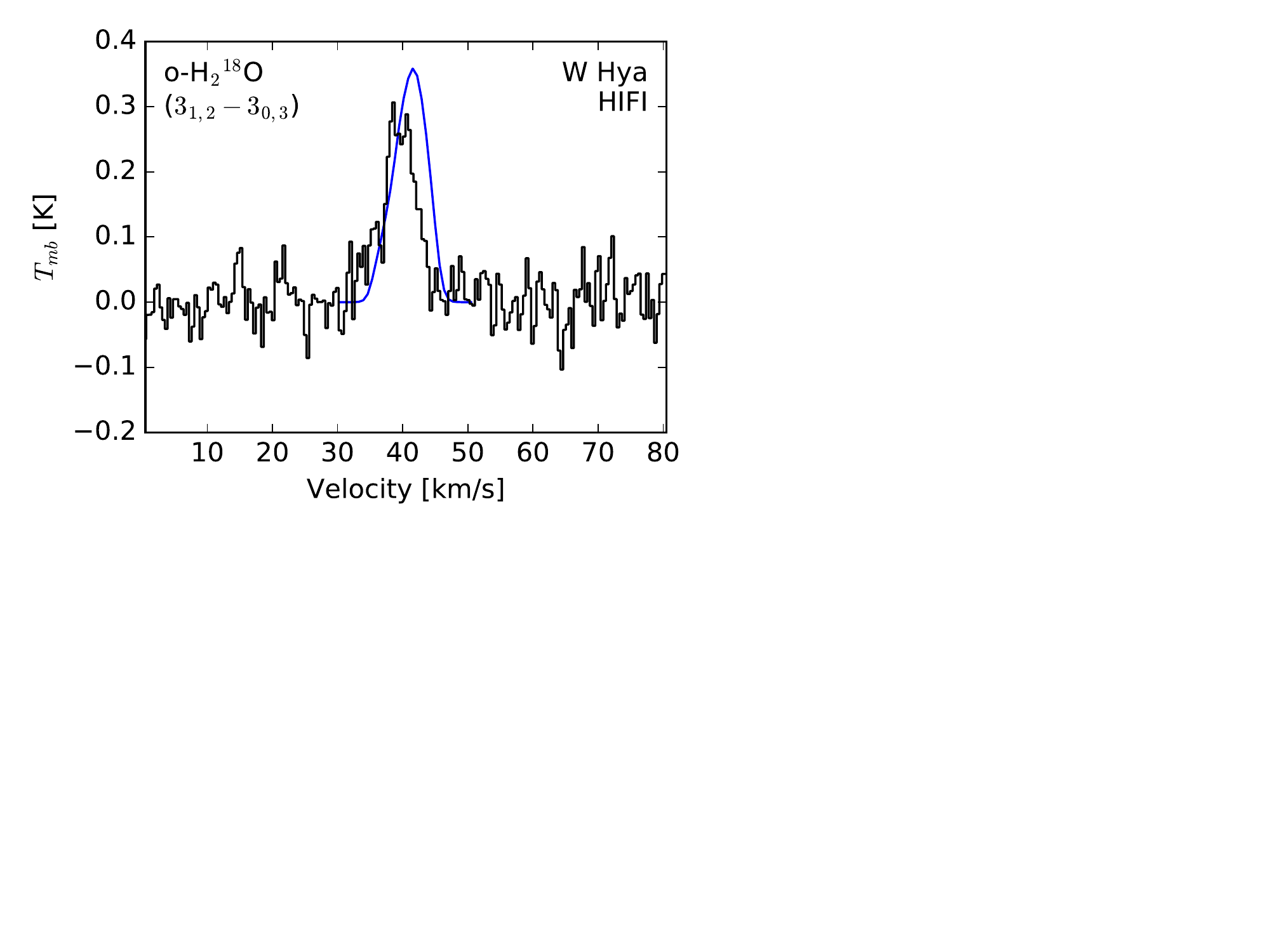}
\includegraphics[width=0.242\textwidth]{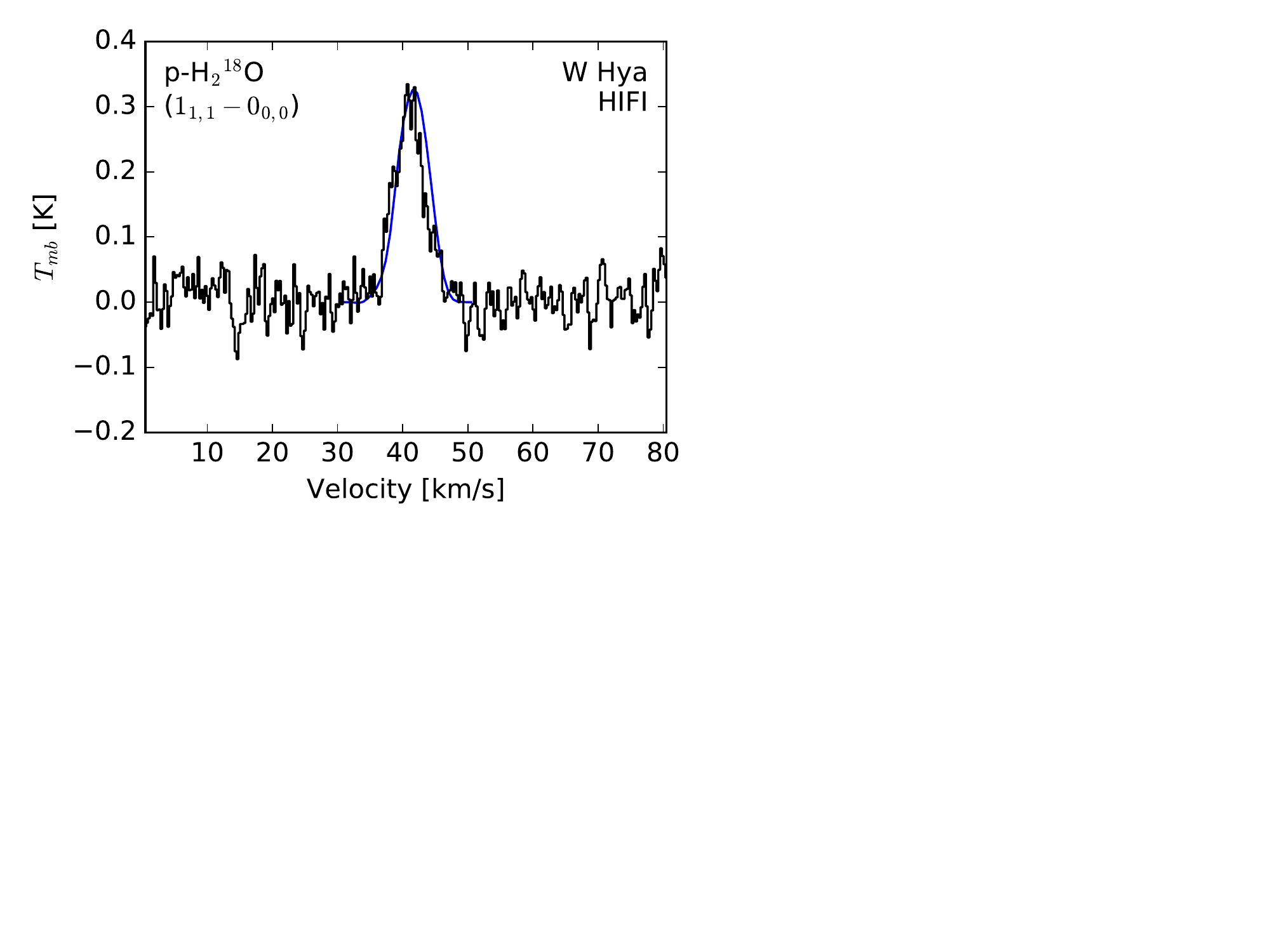}
\caption{HIFI lines (black histograms) and models (blue curves) for W Hya.}
\label{whyaplots}
\end{center}
\end{figure}

\begin{figure}[t]
\begin{center}
\includegraphics[width=0.5\textwidth]{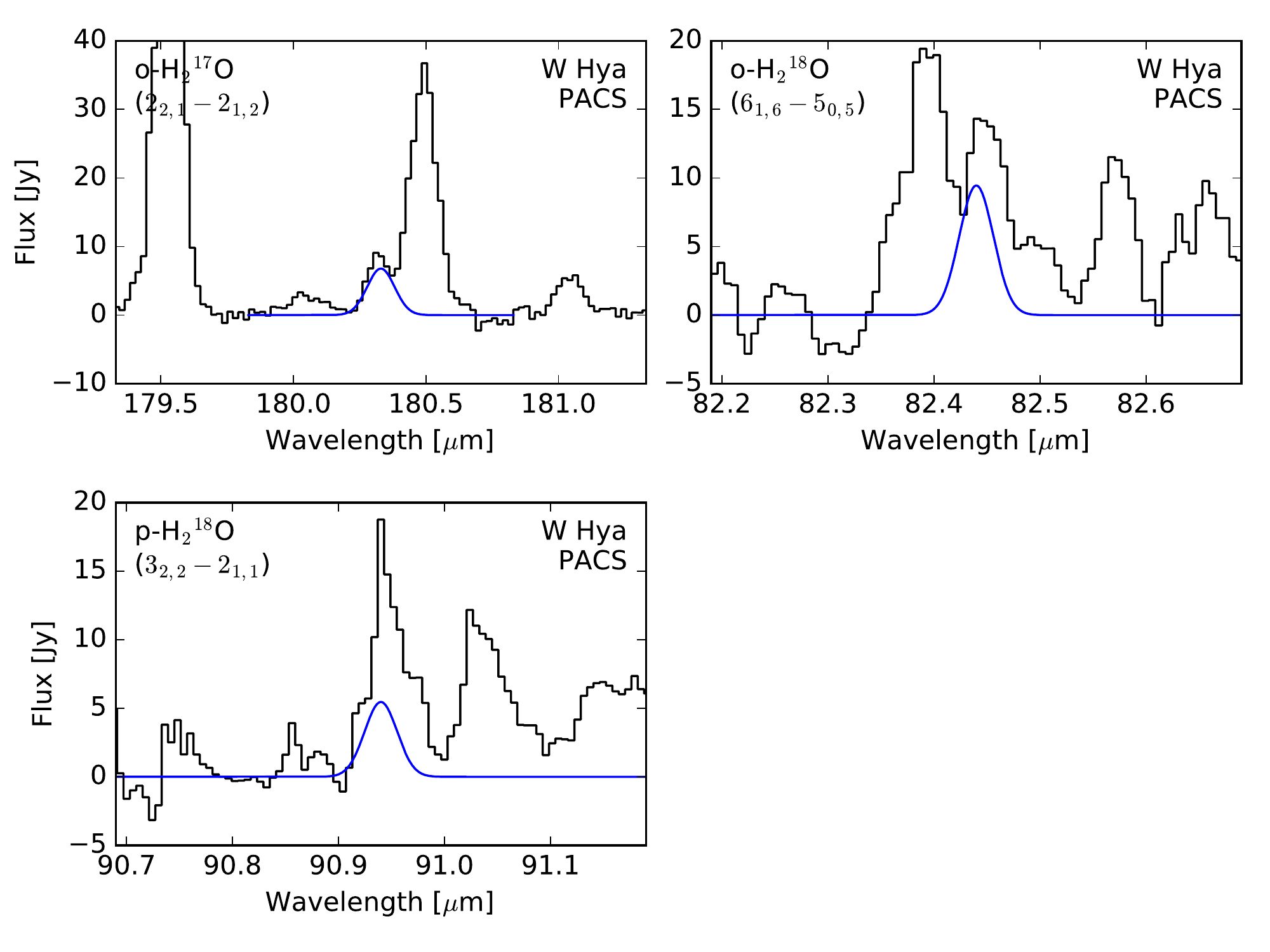}
\caption{ PACS lines (black histograms) and models (blue curves) for W Hya.}
\label{whyapacs}
\end{center}
\end{figure}

\begin{figure}[t]
\begin{center}
\includegraphics[width=0.242\textwidth]{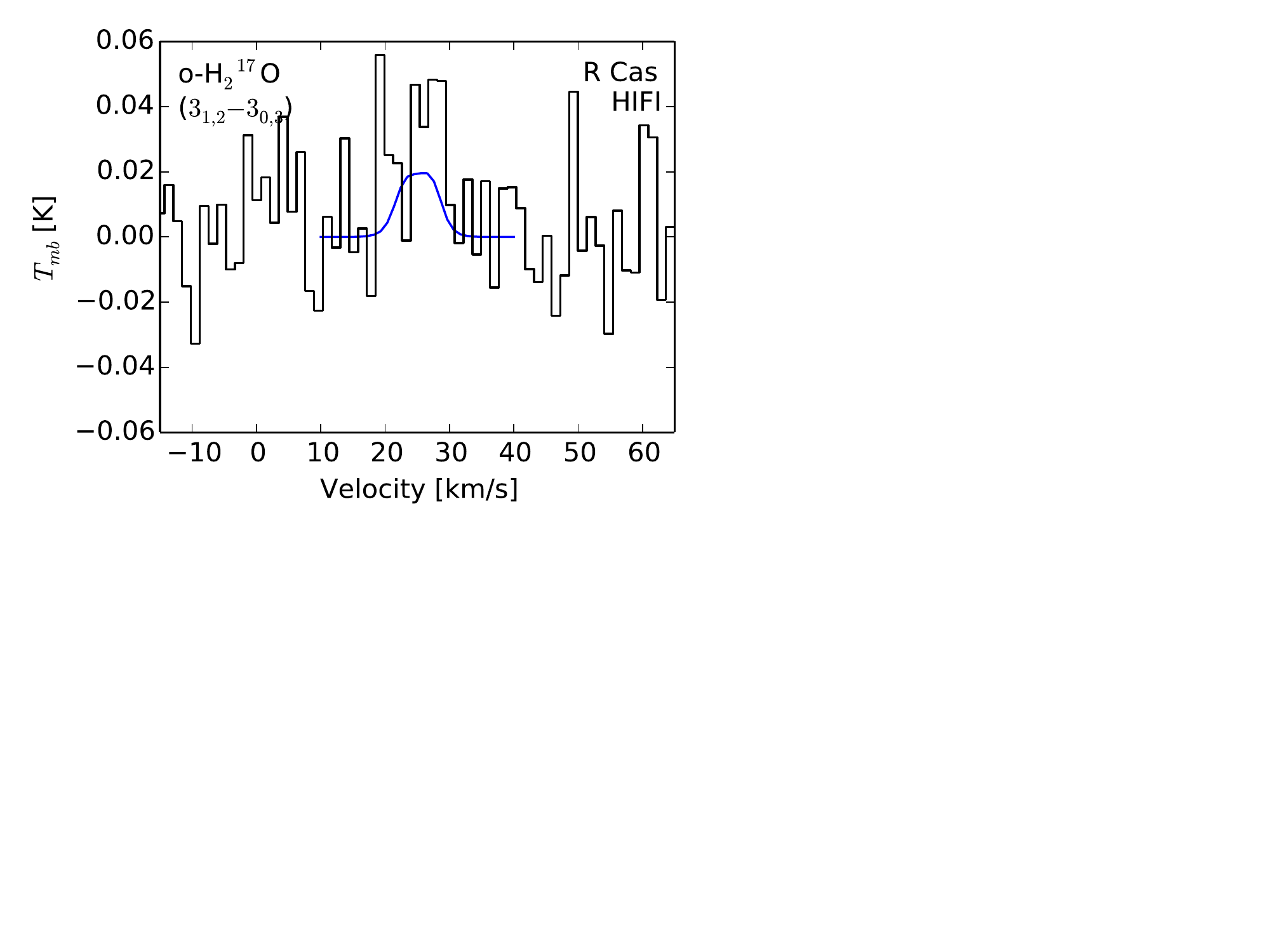}
\includegraphics[width=0.242\textwidth]{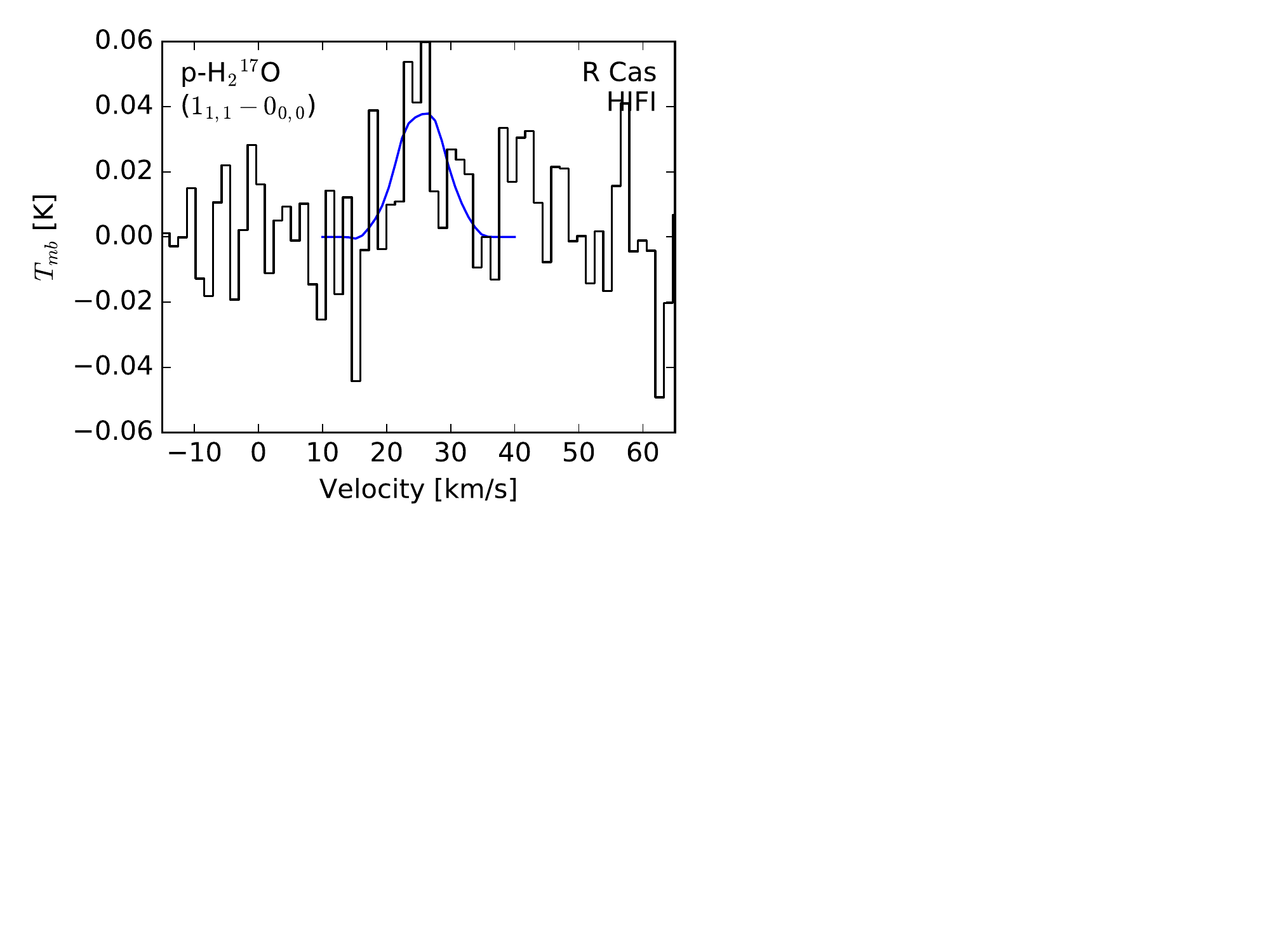}
\includegraphics[width=0.242\textwidth]{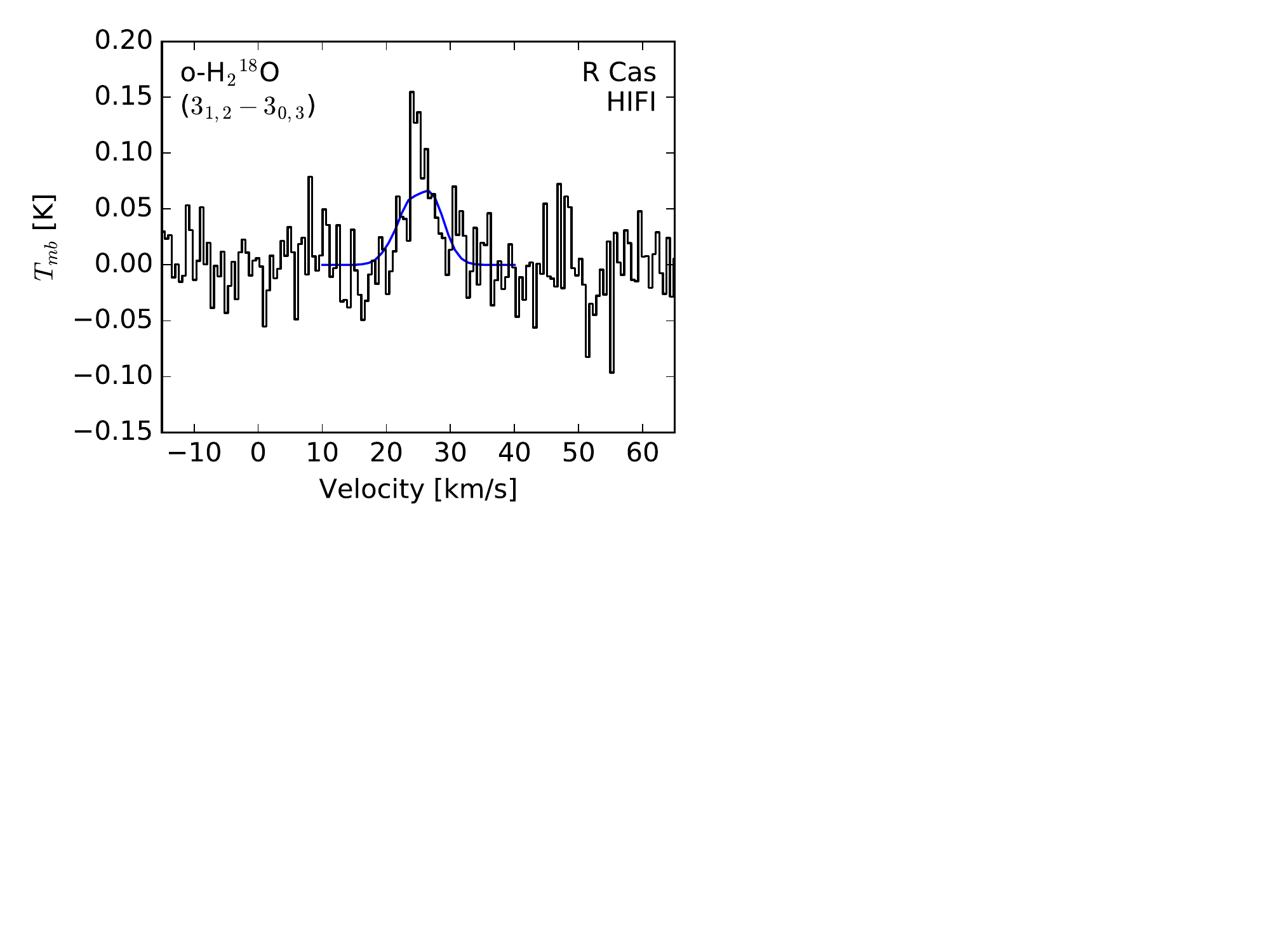}
\includegraphics[width=0.242\textwidth]{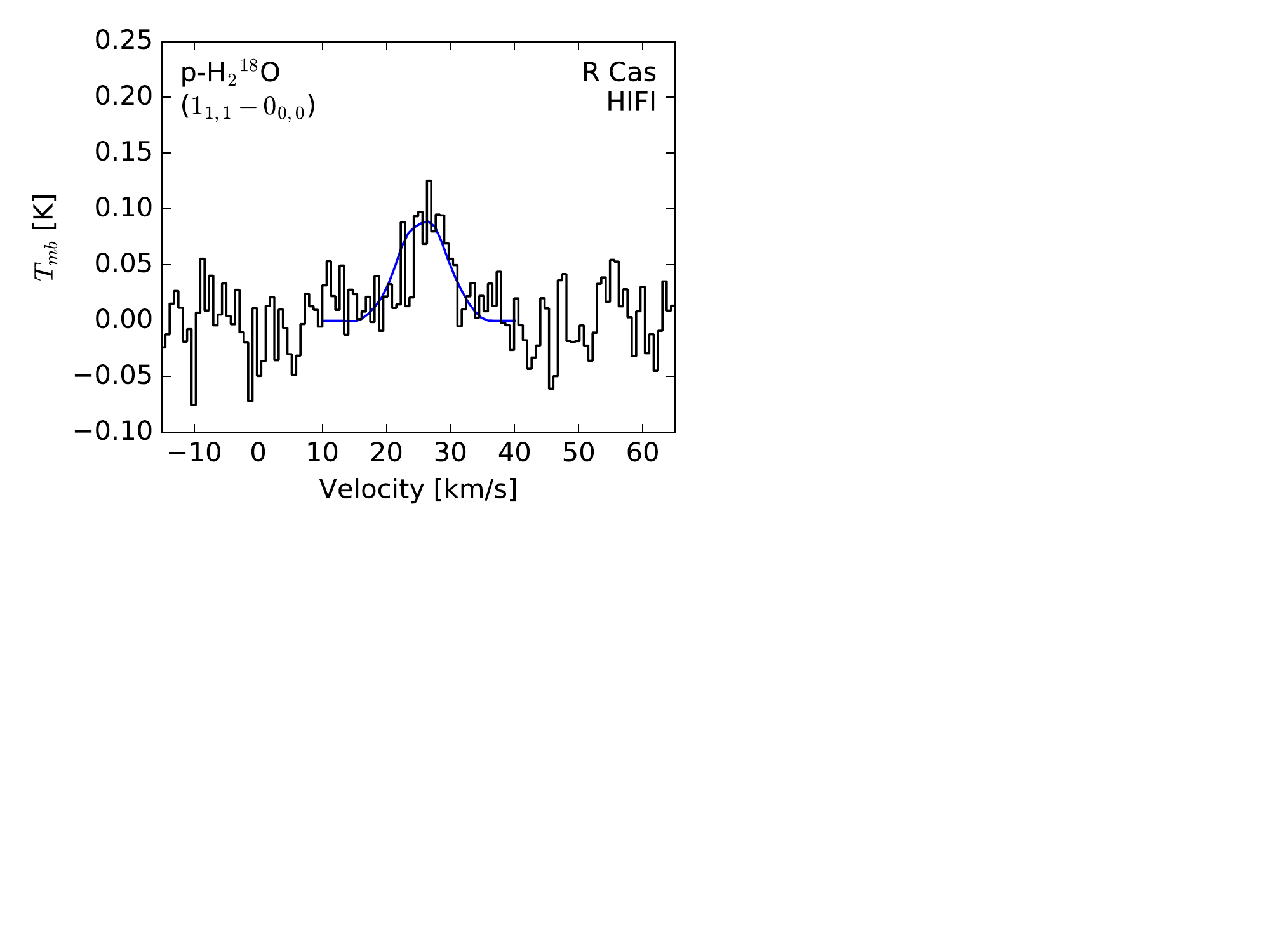}
\caption{HIFI lines (black histograms) and models (blue curves) for R Cas.}
\label{rcasplots}
\end{center}
\end{figure}

\begin{figure}[t]
\begin{center}
\includegraphics[width=0.5\textwidth]{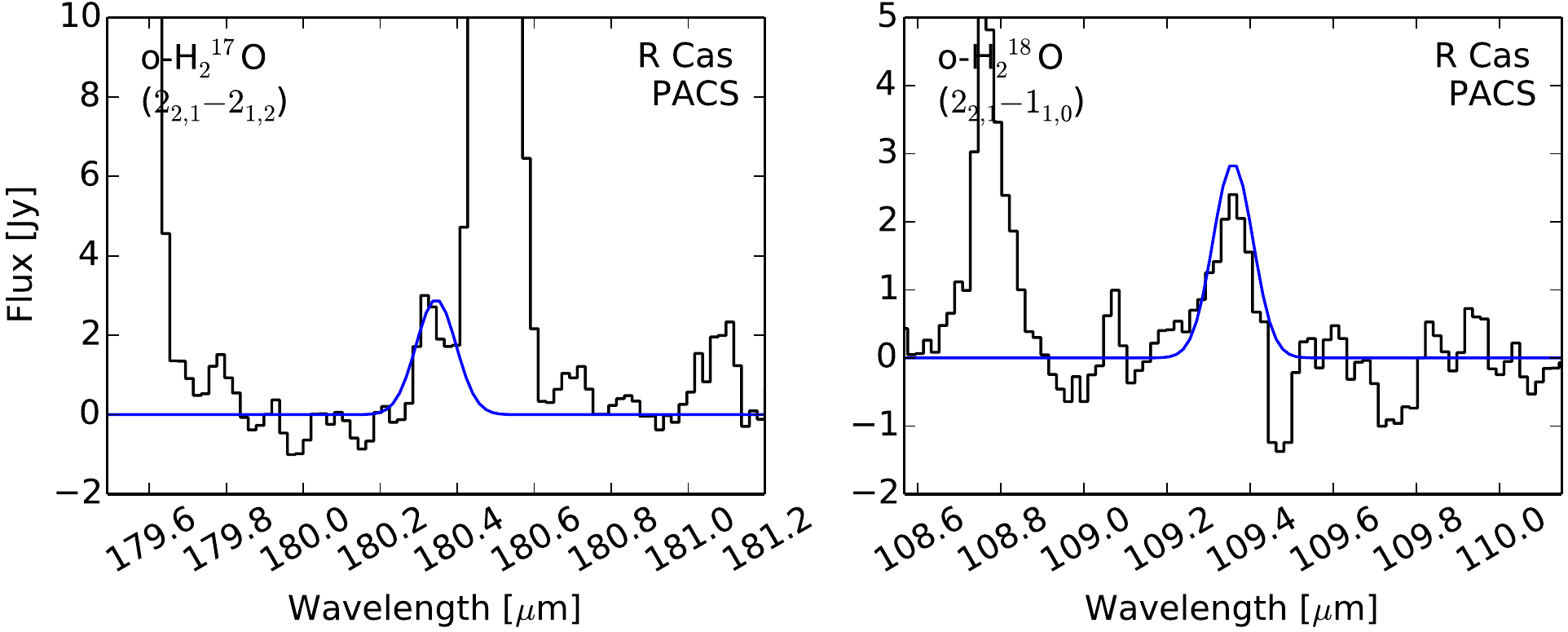}
\caption{ PACS lines (black histograms) and models (blue curves) for R Cas.}
\label{rcaspacs}
\end{center}
\end{figure}

\begin{figure}[t]
\begin{center}
\includegraphics[width=0.242\textwidth]{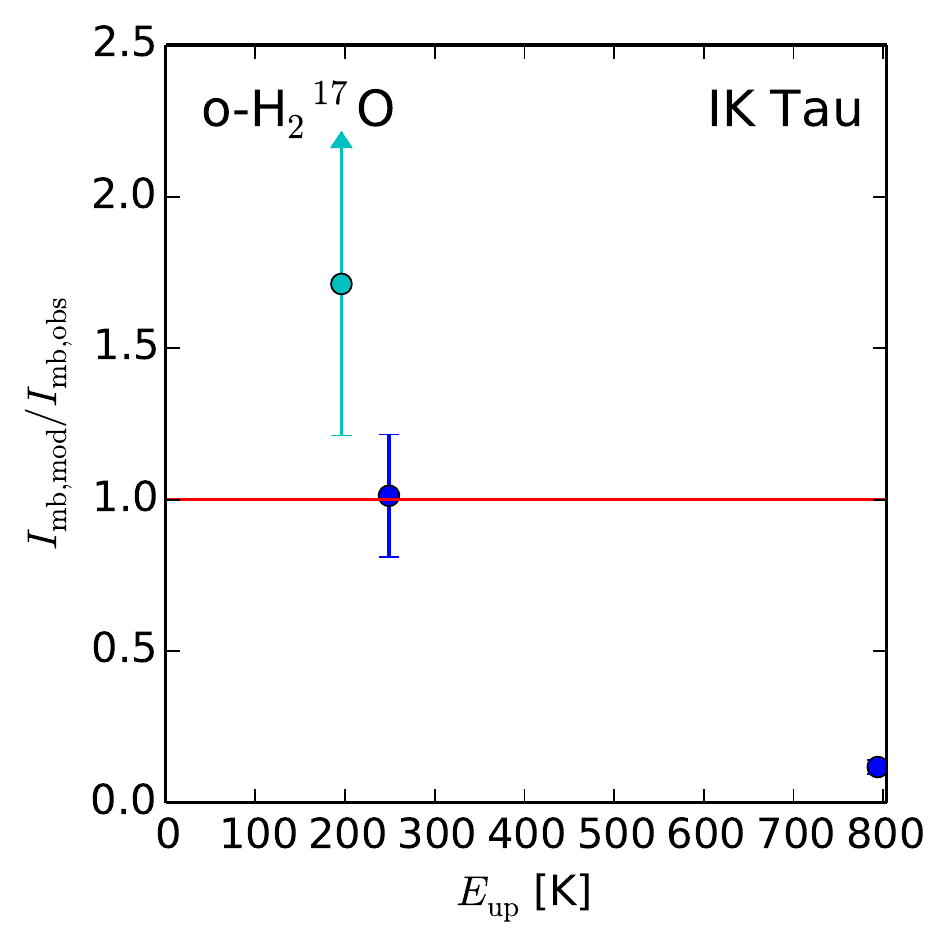}
\includegraphics[width=0.242\textwidth]{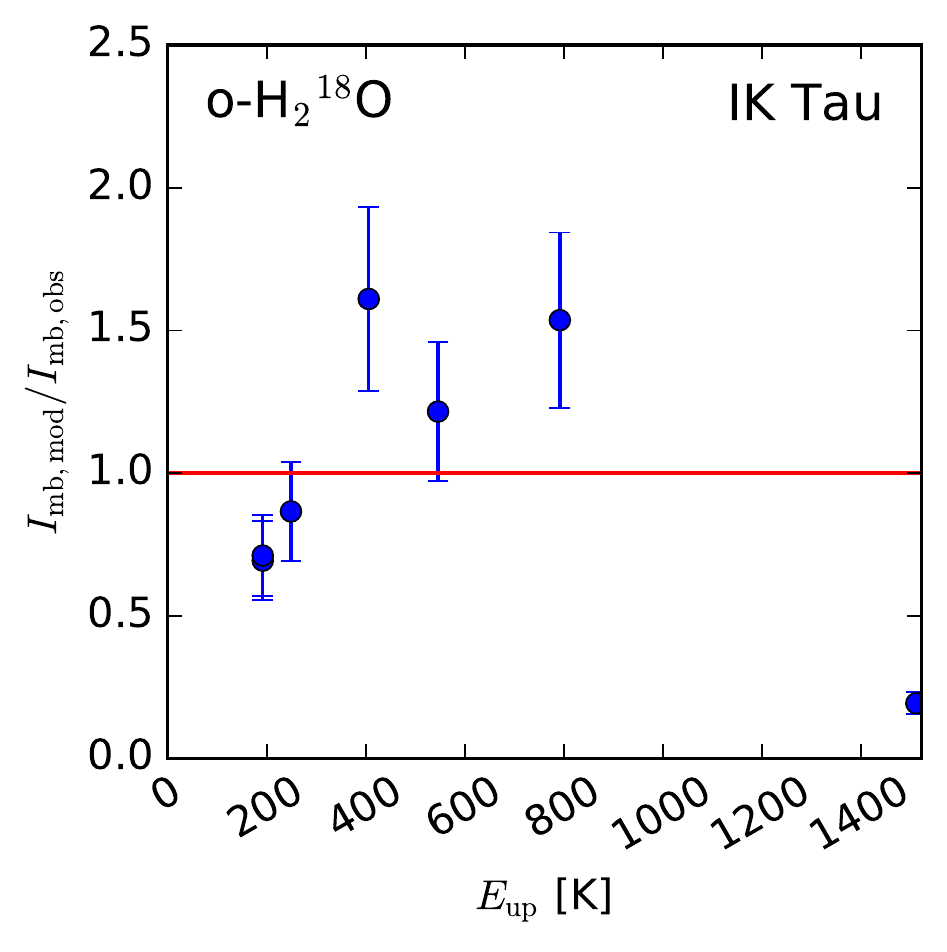}
\includegraphics[width=0.242\textwidth]{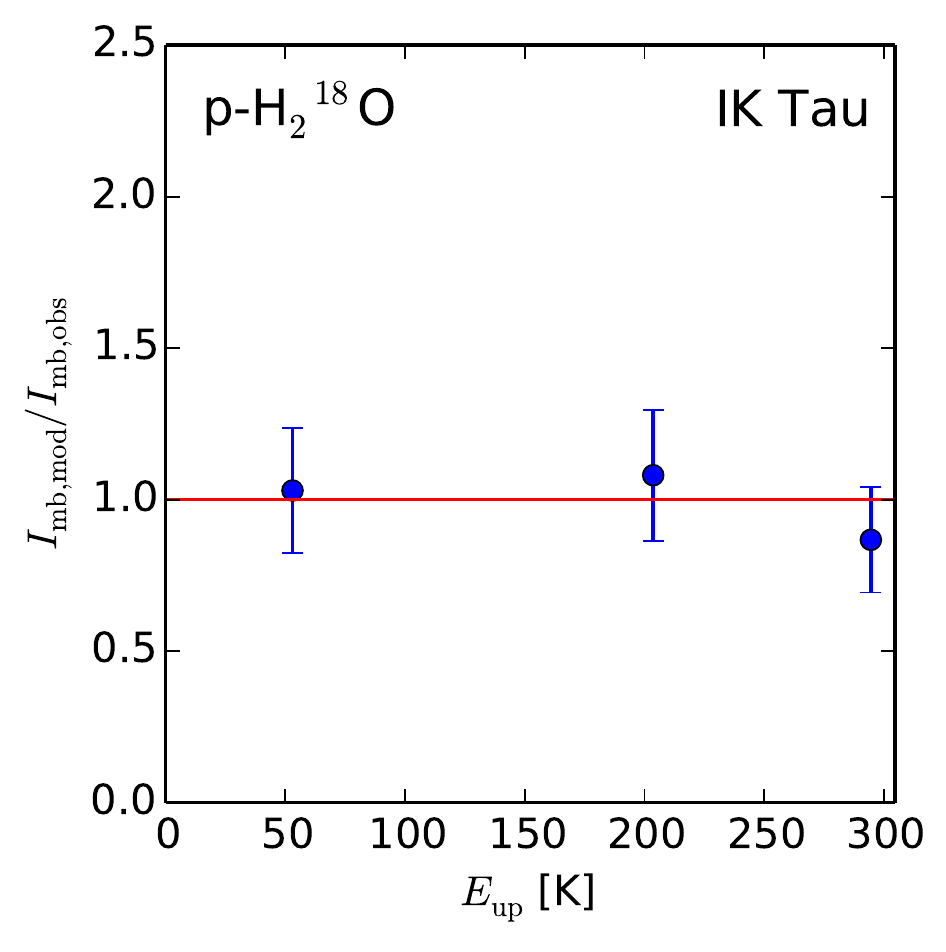}
\includegraphics[width=0.242\textwidth]{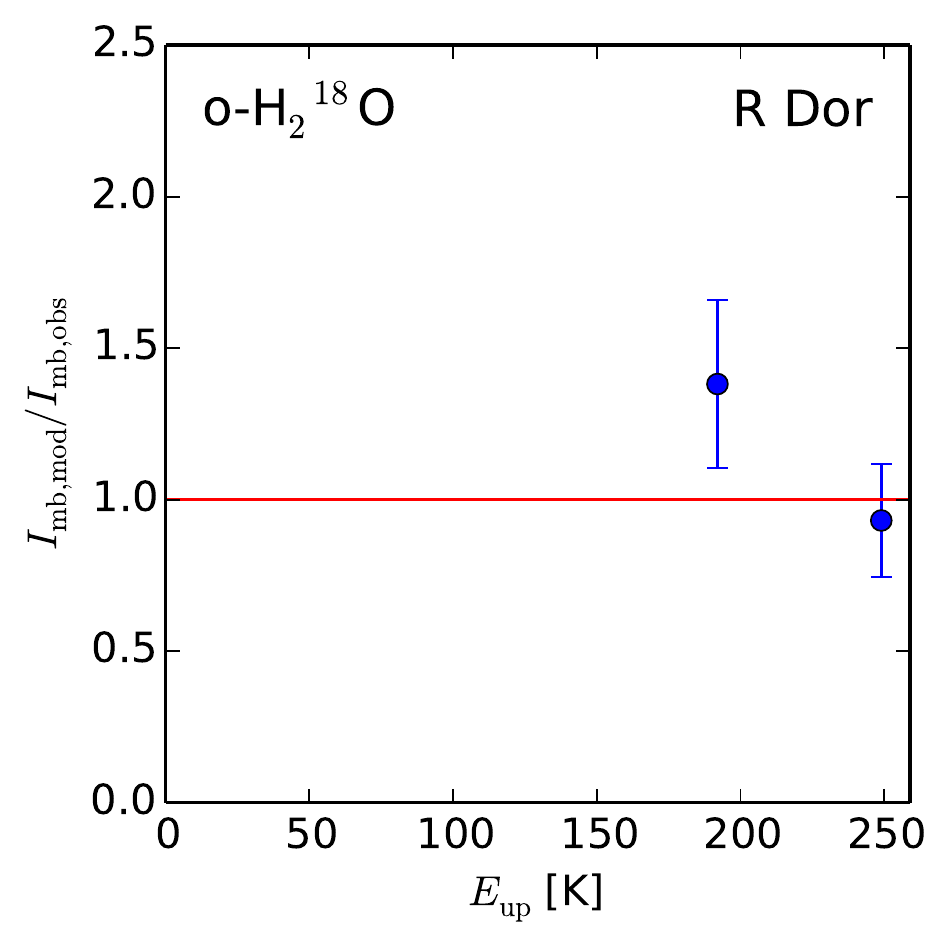}
\includegraphics[width=0.242\textwidth]{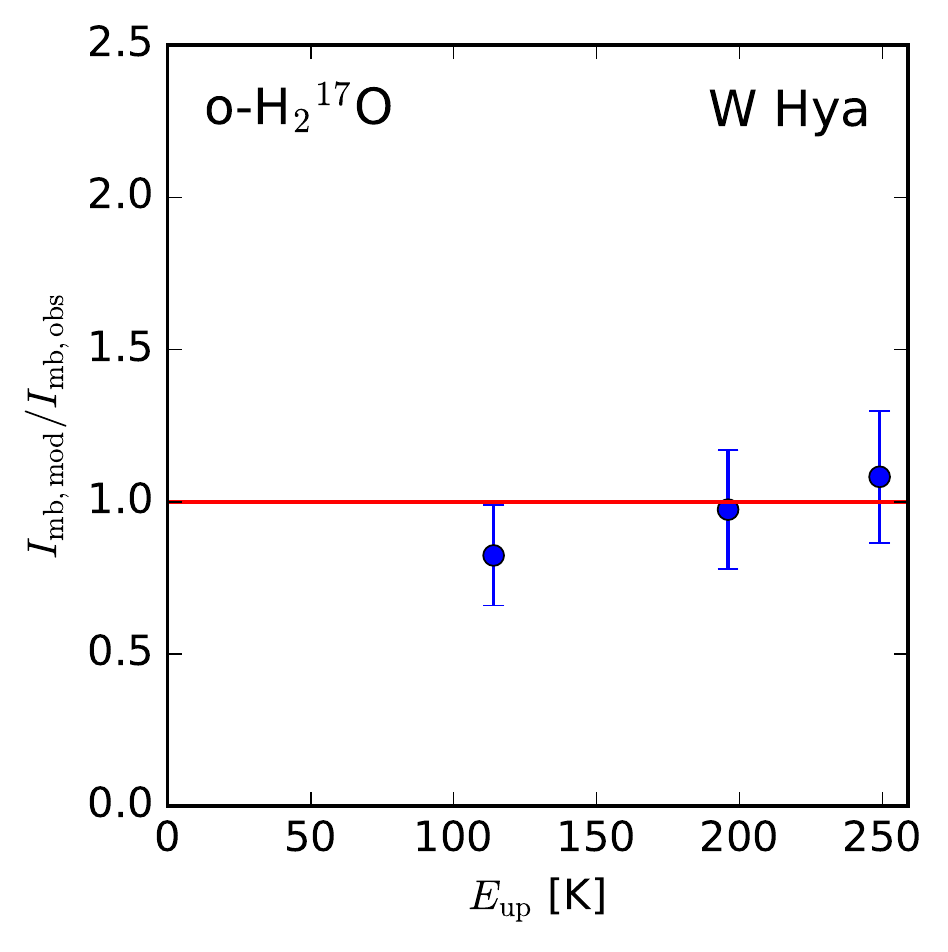}
\includegraphics[width=0.242\textwidth]{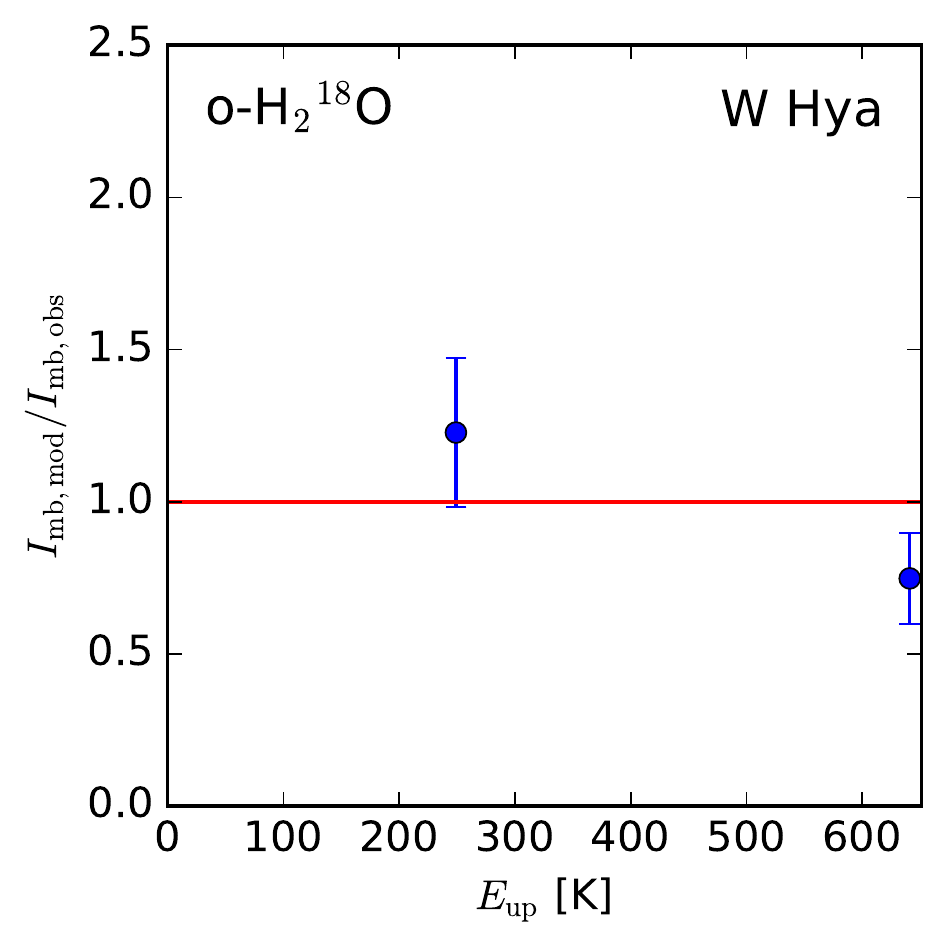}
\includegraphics[width=0.242\textwidth]{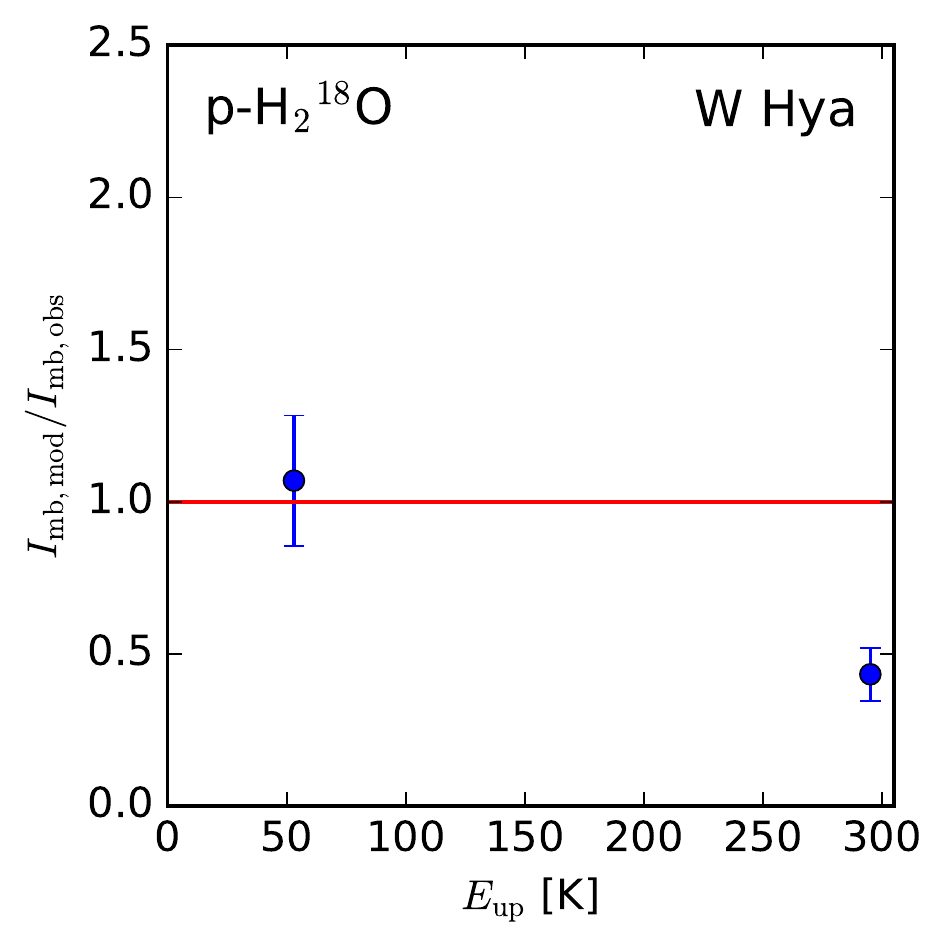}
\includegraphics[width=0.242\textwidth]{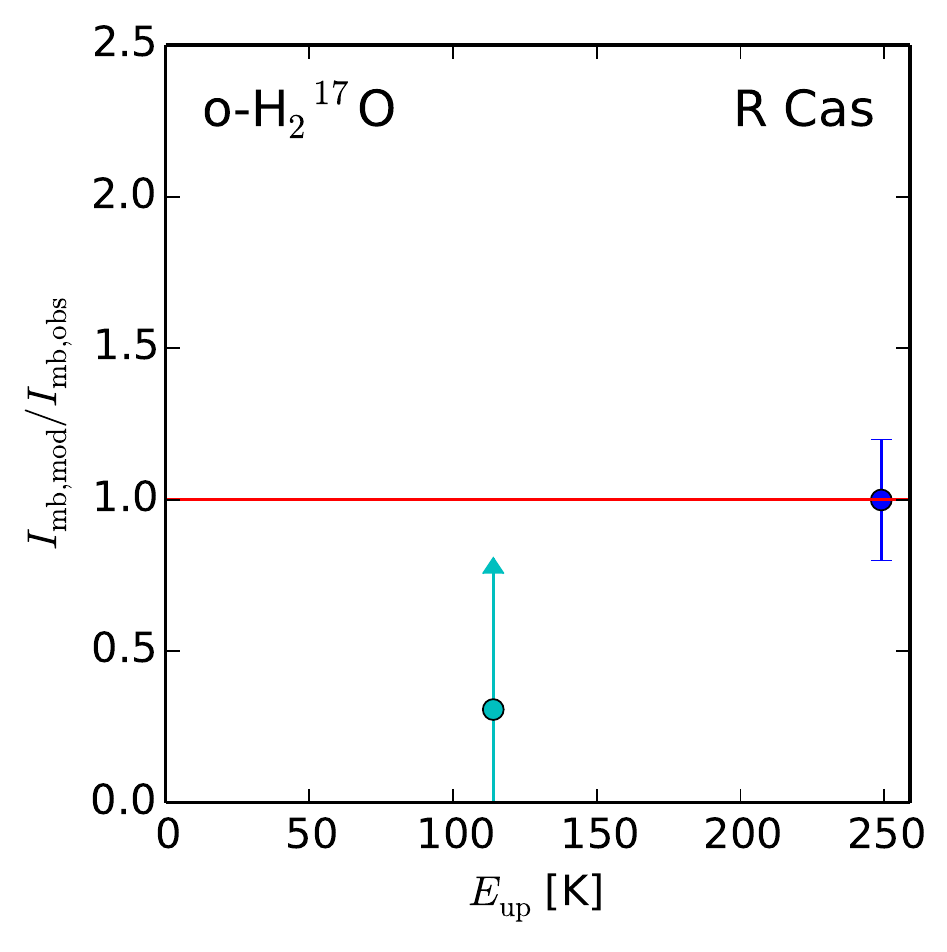}
\includegraphics[width=0.242\textwidth]{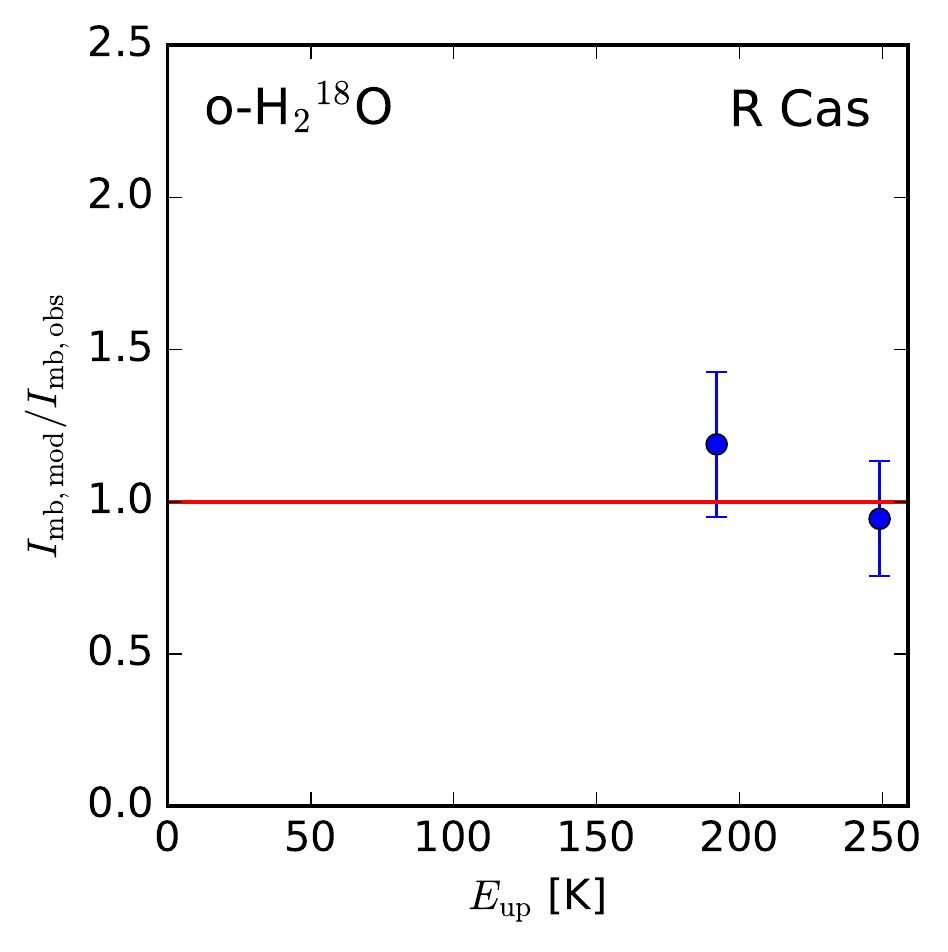}
\caption{Plots indicating goodness of fit for molecules with multiple detected lines.}
\label{fitplots}
\end{center}
\end{figure}

\section{Observation identifiers}

The observation identifiers (ObsIDs) for the \textsl{Herschel} observations used in this study are given in Table \ref{obsids}.

\begin{table}[t]
\caption{ObsIDs for HIFI and PACS observations.}\label{obsids}
\begin{center}
\begin{tabular}{cccc}
\hline\hline
IK~Tau & R Dor & W Hya & R Cas\\
\hline
 1342191651 & 1342197982 & 1342200998 &1342197979\\
1342191650 & 1342197983 & 1342200999 & 1342197978\\
 1342191768& & 1342201788\\
\hline
1342203681 & 1342197795 & 1342212604 & 1342212577\\
1342203680 & 1342197794 & 1342223808 & 1342212576\\
\hline
\end{tabular}
\end{center}
\tablefoot{The upper section contains HIFI ObsIDs and the lower section contains PACS ObsIDs.}
\end{table}%

\end{document}